\documentclass[aps,floatfix,epsfig]{revtex4}
\usepackage{pdfpages}
\usepackage{subfigure}
\usepackage{amsmath}
\usepackage{eqnarray,amsmath}
\usepackage{amssymb}
\usepackage{amsthm}
\usepackage{epstopdf}
\usepackage{mathrsfs}
\usepackage{natbib}
\usepackage{amssymb,latexsym}
\usepackage{dcolumn}
\usepackage{graphicx}
\usepackage{multirow}
\def\be{\begin{equation}}
	\def\ee{\end{equation}}
\def\ba{\begin{eqnarray}}
	\def\ea{\end{eqnarray}}

\begin{document}
\title{Super Interacting Dark Sector: An Improvement on Self-Interacting Dark Matter via Scaling Relations of Galaxy Clusters}

\author{Mahdi Naseri}
\email{mahdi.naseri@email.kntu.ac.ir}
\affiliation{Department of Physics, K. N. Toosi University of Technology, P. O. Box 15875-4416, Tehran, Iran}

\author{Javad T. Firouzjaee}
\email{firouzjaee@kntu.ac.ir}
\affiliation{Department of Physics, K. N. Toosi University of Technology, P. O. Box 15875-4416, Tehran, Iran }
\affiliation{ School of physics, Institute for Research in Fundamental Sciences (IPM), P. O. Box 19395-5531, Tehran, Iran }

\begin{abstract}

\textbf{Abstract:} Self-interacting dark matter is known as one of the most appropriate candidates for dark matter. Due to its excellent success in removing many astrophysical problems, particularly in small scale structure, studying this model has taken on added significance. In this paper, we focus on the results of two previously performed simulations of cluster sized halos with self-interacting dark matter and introduce a new function for the density profile of galaxy clusters, which can perfectly describe the result of these simulations. This density profile helps to find a velocity dispersion profile and also a relation between cluster mass and concentration parameter. Using these relations, we investigate two scaling relations of galaxy clusters, namely mass-velocity dispersion and mass-temperature relations. The scaling relations reveal that in the self-interacting dark matter model, halos are more massive than  what the standard non-interacting model predicts for any fixed temperature. We also study the mass-temperature relation for a hybrid interacting model, which is a combination of self-interacting dark matter  idea with another model of the dark sector in which dark matter particle mass is determined according to its interaction with dark energy. This super interacting dark sector (SIDS) model can change the mass-temperature relation to  a modified form that has the same result as a non-interacting model. Finally, we provide quantitative expressions which can describe the constants of this interacting model with the value of cross-section per unit mass of dark matter particles.  

\end{abstract}
%
%

\maketitle

\tableofcontents
\section{Introduction}

Dark matter (DM) is believed to be the second largest component of the total mass-energy of the Universe, accounting for about five times as much as the contribution of baryonic matter \cite{Ade:2015xua}. Studying velocity dispersion of galaxies in the Coma cluster \cite{Zwicky:1937zza}, rotation curves of the Andromeda galaxy \cite{Babcock}, and spiral galaxies \cite{Rubin} were the preliminary reasons that led to the idea of DM. In more recent efforts, many other cosmological and astrophysical tests have verified the existence of DM in the cosmos, including x-ray observations of the Coma cluster \cite{Briel:1997hz}, the Bullet cluster observations \cite{Komatsu}, gravitational lensing of galaxies \cite{AdelmanMcCarthy:2005se} and cosmic microwave background (CMB) anisotropies \cite{Challinor:2012ws}. \\

Despite strong evidence for the existence of DM, its nature has been under discussion for a prolonged period of time. There are a wide range of candidates that have been proposed in particle physics to justify different cosmological effects of DM, and no absolute consensus has been built yet among physicists. Some of those candidates had been under study for a while but were finally ruled out by further exploration. For example, once stellar remnants (such as black holes and neutron stars) were considered as DM candidate, namely massive compact halo objects (MACHOs), which can explain the missing mass of the galaxies \cite{Mohapatra:1999ih}. However, it was found that MACHOs can only contain about $3\%$ of the Milky Way dark matter mass \cite{Bahcall:1994xu,Freese:2000vi}. \\ 

On the contrary, many candidates in particle physics can still attract attention as a DM model, such as Axions and WIMPs, to name but a few. Axion was postulated as a Peccei-Quinn solution to the strong-CP problem in quantum chromodynamics \cite{Peccei:1977ur} and immediately considered as a DM candidate \cite{Weinberg:1977ma}. Axionic DM particles are neutral low-mass particles that can interact via the weak and the strong forces and decay into microwave photons. Thus, Axionic DM could theoretically be detected with the aid of these emitted photons. The other introduced candidate, WIMPs (weakly interacting massive particles), might be the most interesting suggestion. WIMPs merely interact via the weak force (no interaction through the electromagnetic force justifies no emission of light) and their mass ranges between 2 GeV (Lee-Weinberg bound) and 100 TeV \cite{Roszkowski:2017nbc}. Given these properties, one can theoretically obtain the relic density of WIMPs consistent with its right and observed value in the Universe. This amazing feature is known as the 'WIMP miracle' in the literature. \\ 

The standard $\Lambda$CDM model with WIMPs has been very successful in describing large scale structures. Nevertheless, there are a number of problems with this model in small scales of the cosmos. The first problem stems from the detection experiments of DM. These experiments try to detect DM particles via thair scattering with atomic nuclei in underground laboratories. While 'SuperCDMS' \cite{Agnese:2017njq} and 'PICO' \cite{Amole:2017dex} experiments found the lower limit of DM-nucleon scattering cross-section to be somewhere in the region of $10^{-42}-10^{-45} \; \rm{cm}^2$ for DM particle with the mass of $\approx$ 10 GeV, theoretical predictions attribute this value to WIMPs with the mass of $\approx$ 100 GeV. This discrepancy between theory and expriment also remains manifest in some other expriments, showing an even larger gap between theory and data. In addition, cold DM (CDM) cosmological simulations predict the formation of $\approx$ 500 subhalos around the Milky Way \cite{Moore:1999nt}, whereas solely about $\approx$ 50 satellites and dwarf spheroidal galaxies (dSphs) have been identified around the Milky Way so far \cite{satellite}. These simulations \cite{Springel:2008cc} also show that hundreds of thousands of satellites should be formed within the Milky Way, with masses of $\approx 10^5 \; \rm{M}_\odot$ or larger. Since the number of predicted satellites for galaxy sized halos is way higher than discovered satellites, this contradiction has been known as the 'missing satellite problem'. Eventually, the so-called 'core-cusp problem' is another challenge for collisionless DM. Although CDM simulations of the density profile of halos, embracing Navarro-Frenk-White (NFW) density profile \cite{Navarro:1996gj}, lead to a steep slope of density profile in the central region of halos (cuspy profile), galactic rotation curves do not indicate a steep slope and instead, reveal rather cored profiles near the center \cite{cored}. \\

All these mentioned cosmological problems can be solved considering a type of collisional DM with a non-negligible cross-section. The idea of self-interacting dark matter (SIDM) was initially proposed \cite{Spergel:1999mh} in order to resolve these problems and rapidly turned into an interesting alternative for CDM. In this model, DM particles can interact with themselves as they have a non-negligible cross-section, similar to that of the strong force. It has been proved that such an assumption may result in a heat transfer that reduces the density of the central region of halos and consequently, eliminates the core-cusp problem. The decline of density in the inner region of halos also leads to the formation of fewer satellites and removes the missing satellite problem. \\

Particle physics has a number of candidates for SIDM, including Q-balls \cite{Kusenko:2001vu}, hidden sectors in String Theory \cite{Faraggi:2000pv} and hidden charged dark matter \cite{Feng:2009mn}. Basically, SIDM may comprise two different possibilities with regard to its scattering cross-section per unit mass, i.e. $\frac{\sigma}{m}$. In the first possibility, self-interaction results from assuming hard-sphere scattering among DM particles, which suggests a constant value of $\frac{\sigma}{m}$. On the other hand, $\frac{\sigma}{m}$ could be velocity dependent considering a Yukawa potential for SIDM scattering in the second possibility \cite{Loeb:2010gj}. \\

There are also different constraints on the value of a constant $\frac{\sigma}{m}$ which have been obtained so far. A lower limit of approximately $\frac{\sigma}{m}>0.1 \; \rm{cm}^2 g^{-1}$ is required for SIDM to be distinguishable from CDM on small scales \cite{Zavala}. The shape of clusters can be studied to find an upper limit on velocity independent cross-section per unit mass of SIDM. In this sense, the value of nearly $\frac{\sigma}{m}<1 \; \rm{cm}^2 g^{-1}$ is obtained \cite{Peter:2012jh}. In fact, most physical and observational constraints have been derived in this range and it would be a suitable choice to consider $0.1 \; \rm{cm}^2 g^{-1}<\frac{\sigma}{m}<1 \; \rm{cm}^2 g^{-1}$ for acceptable cases. \\

Galaxy clusters could be as much helpful as small scale structures to conduct research on SIDM. In this regard, simulations play a significant role, and comparison between their results and observations provide a host of opportunities to investigate cluster sized halos of SIDM. In this work, we use the outcomes of previously performed simulations, in order to study halo profiles and scaling relations of galaxy clusters, and ultimately introduce a novel description of SIDM through connecting our results to another dark energy (DE) model. \\ 

The paper is organized as follows: Section II is dedicated to finding quantitative relations for halo profiles, embracing density profile and velocity dispersion profile, according to the result of simulations. A mass-concentration relation can also be derived regarding the modified density profile. In Section III, two important scaling relations are obtained, namely mass-velocity dispersion and mass-temperature relations. As the obtained mass-temperature relation suggests higher masses for given temperatures compared with the non-interacting case, an improved model is investigated in Section IV which can reform the SIDM model and removes this difference. Finally, we conclude and summarize the main results in Section V. \\

\section{Halo Profiles}

Characterizing a halo includes various profiles and variables, such as density profile, total mass, concentration parameter, halo shape and velocity dispersion profile. The assumption of self-interaction among DM particles may dramatically alter these characteristics in both large and small scale structures, and the imprint of collisional DM can emerge in a variety of astrophysical tests. For example, the effect of SIDM on galaxy wraps is investigated in Ref. \cite{Pardo:2019wie}. \\

In order to study SIDM cluster sized halo profiles, Brinckmann et al. performed a dark-matter-only simulation, containing 28 dynamically relaxed halos with mass of $\approx 10^{15} \; \rm{M}_{\odot}$ in CDM and SIDM models \cite{Brinckmann:2017uve}. In a more comprehensive and more recent work, Andrew Robertson et al. performed another similar simulation, but with adding baryonic matter effects and also another mass window of $\approx 10^{14} \; \rm{M}_{\odot}$ \cite{Robertson:2018anx}. Both studies have been performed for two choices of $\frac{\sigma}{m}$, which are $\frac{\sigma}{m}=0.1 \; \rm{cm}^2 g^{-1}$ (hereafter SIDM0.1) and $\frac{\sigma}{m}=1 \; \rm{cm}^2 g^{-1}$ (hereafter SIDM1). Even though Andrew Robertson et al. have also studied the velocity-dependent model of $\frac{\sigma}{m}$ as a part of their simulation, this model is not our objective in this paper. We concentrate on the results of these simulations only for the constant  $\frac{\sigma}{m}$ model (mainly the one in Ref. \cite{Robertson:2018anx}), and try to convert those results to mathematical expressions that can be used to determine halo characteristics and to derive scaling relations in the next Section. As we deduce our relations from \cite{Robertson:2018anx}, we can solely obtain and use our equations for $0.1 \; \rm{cm}^2 g^{-1}<\frac{\sigma}{m}<1 \; \rm{cm}^2 g^{-1}$ and $10^{14}<\frac{\rm{M}}{\rm{M}_{\odot}}<10^{15}$, analogous to the upper and lower limits of the given simulation. Meanwhile, all the results are found for low-redshift clusters, due to the assumption of $z \approx 0$ in the simulation.

\subsection{Density Profile}

 The NFW density profile \cite{Navarro:1996gj} is one of the well-known density profiles according to a CDM simulation. This profile is given by

\begin{equation} \label{NFW}
\frac{\rho}{\rho_0}=\frac{1}{x(1+x)^2} \, ,
\end{equation}

where $x=\frac{r}{r_s}$ and $r_s$ is the scale radius of halo. In this profile, $\rho_0=\rho_{crit}\delta_c$ is calculated for each halo, where $\rho_{crit}$ is the critical density of the Universe at given redshift (z) and $\delta_c$ depends on the concentration parameter (c) with the relation of

\begin{equation} \label{delta_c}
\delta_c=\frac{200}{3}\frac{c^3}{ln(1+c)-c/(1+c)} \, .
\end{equation}

For a flat cosmos, the overdensity of a virialized halo ($\Delta_{vir}$) is given by the relation of \cite{overdensity}
\begin{equation} \label{overdensity}
\Delta_{vir}\approx \frac{18 \pi ^2 + 82 \chi - 39 \chi ^2}{\Omega (z)} \, ,
\end{equation}
where $\chi=\Omega (z)-1$, and $\Omega (z)$ is the ratio of mean matter density to the critical density at given redshift. Various measurements of $0.3 \lesssim \Omega (z=0) \lesssim 0.4$ result in different values of $\Delta_{vir}$, which are less than $\approx 337$. At a higher reshift, $\Delta_{vir}$ has smaller value, and for the cosmology of Einstein-deSitter, overdensity of a virialized halo is found to be $\Delta_{vir} \approx 178$ at all time. As a result, it is a common choice in the literature to consider $\Omega \approx 200$ for low-reshift virialized halos. Note that the value of $\Delta_{vir}$ merely depends on redshift. This choice also means that the radius inside which the mean density is equal to $\approx200$ times the critical density of the Universe is the radius of virialized halo ($r_{200} \approx r_{vir}$), and $\rm{M}_{200}$ indicates the enclosed mass by this radius. With this assumption, the relation of $c=\frac{r_{200}}{r_s}\approx \frac{r_{vir}}{r_s}$ connects concentration parameter to the scale radius and $r_{200}$ of a cluster in the NFW profile. \\

\begin{figure}[h]
 	\centering
 	\includegraphics [width=0.7 \linewidth] {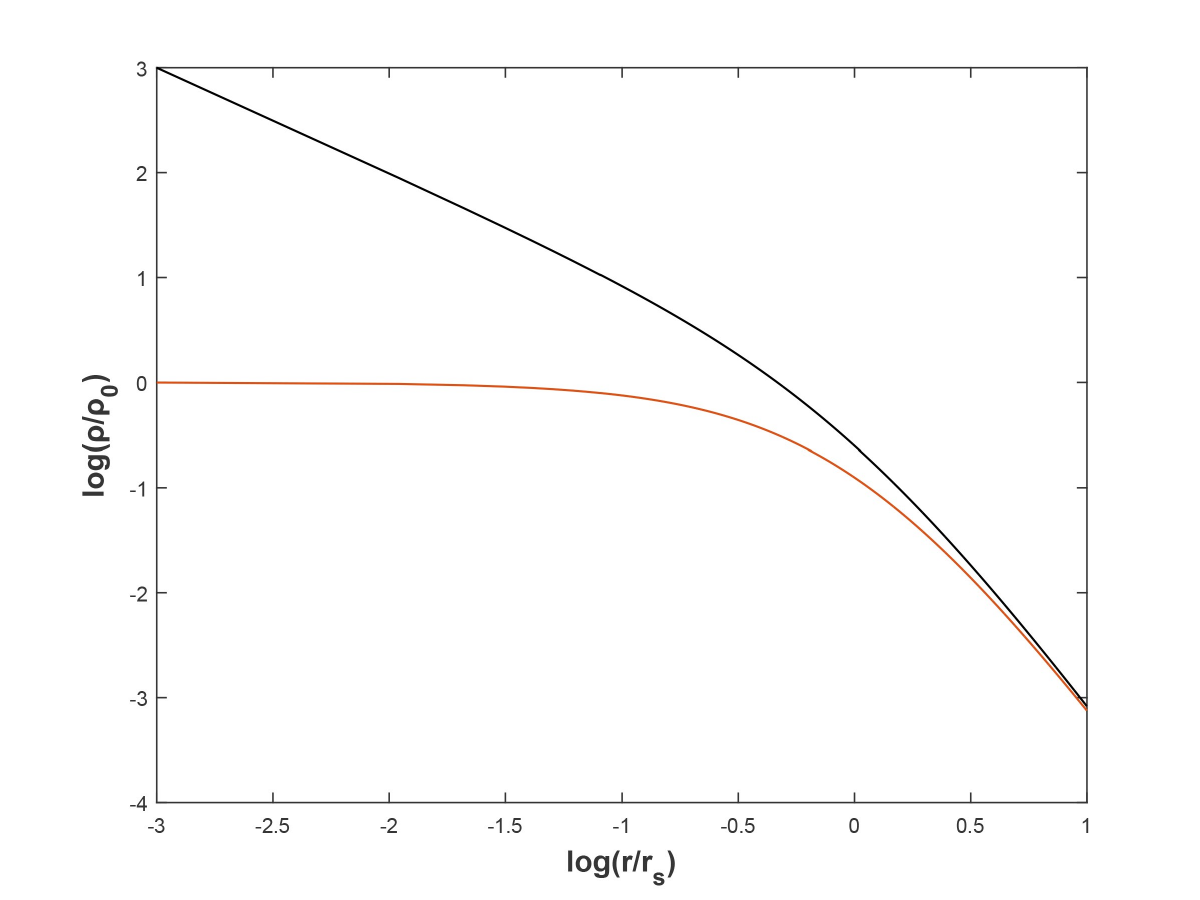}
 	\caption{This figure illustrates the NFW profile (black curve) and Eq.~(\ref{density_profile}) without the second term in its numerator (orange curve), that is $f(x)=\frac{1}{x^{\alpha}(1+x)^{3-\alpha}}$ (with $\alpha = 0$ in this particular example). Such a profile results in an over-large core, which is not compatible with the simulations.}
 	\label{core}
 \end{figure}

\begin{figure}[h]
 	\centering
 	\includegraphics [width=0.7 \linewidth] {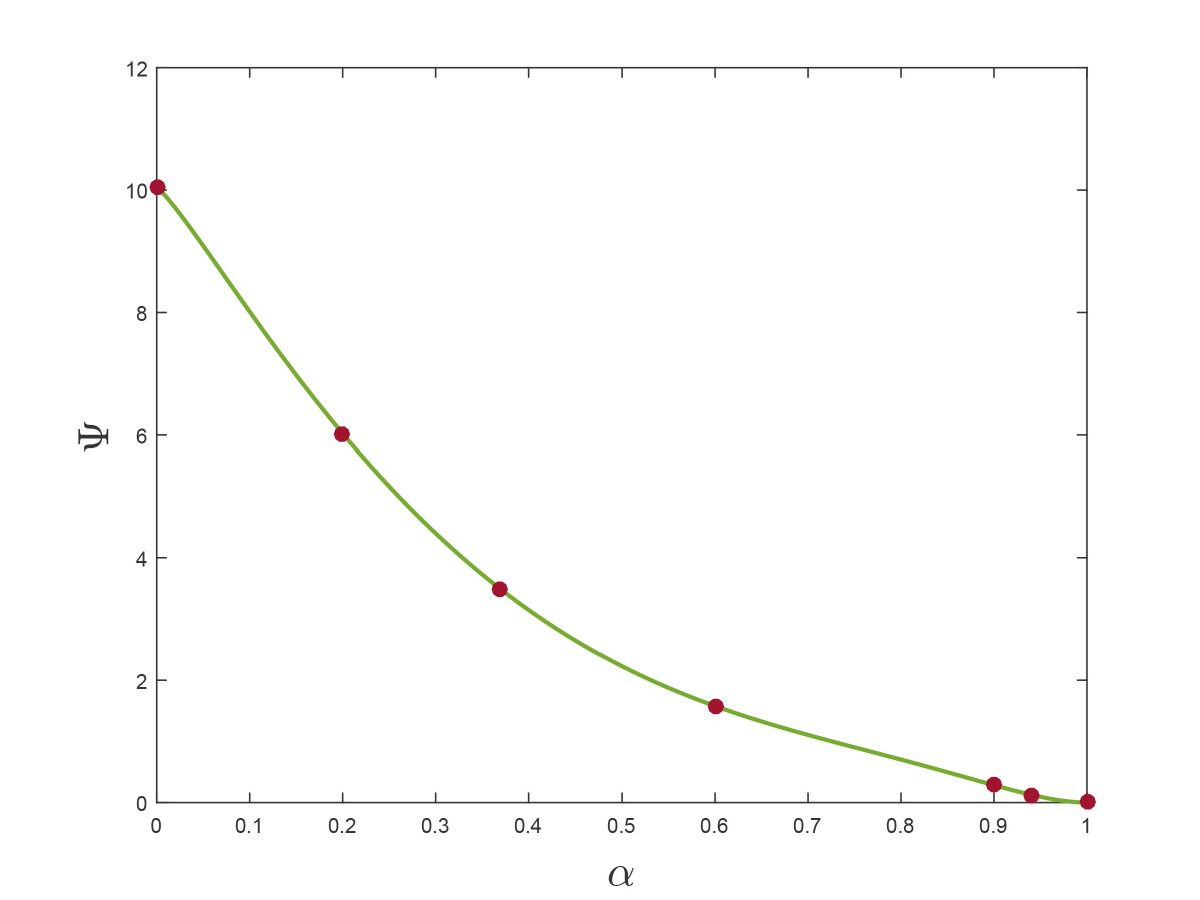}
 	\caption{The behavior of $\Psi$ as a function of $\alpha$. Here, maroon circles show some of the numerical values that have been determined for Eq.~(\ref{density_profile}) to become compatible with simulation results. The green curve is the fitted equation to those values, i.e. Eq.~(\ref{profile_parameters}).}
 	\label{Psi_fig}
 \end{figure}

It can easily be seen from Eq.~(\ref{NFW}) that at small radii ($x\to 0$), NFW density profile reduces to $\rho \propto r^{-1}$, while at large radii ($x\to \infty$), it changes with radius as $\rho \propto r^{-3}$. As has been mentioned before, observation of cored profiles ($\rho =constant$) gave rise to efforts to reform CDM simulations. However, SIDM simulations managed to resolve this problem in galaxy sized halos. Even in larger scales, e.g. clusters of galaxies, SIDM makes a difference. According to simulations (see the top plots of Fig. (2) in Ref. \cite{Robertson:2018anx}), the slope of density profile slowly changes from $\rho \propto r^{-1}$ (consistent with NFW profile) to $\rho =constant$ at small radii with two parameters: the total mass of the cluster and the value of $\frac{\sigma}{m}$. The more massive the cluster is, the less cuspy the profile is. Likewise but more effectively, larger $\frac{\sigma}{m}$ leads to a more apparent cored profile. With respect to this behavior, we propose a mathematical function that describes this change and has the form of
\begin{equation} \label{density_profile}
f(x)=\frac{\rho (x)}{\rho_0}=\frac{1+\Psi(1+x)^{-2}e^{-x/k}}{x^{\alpha}(1+x)^{3-\alpha}} \, .
\end{equation}
To find this density profile from the simulation, we started from the NFW profile and added several modifications to that in order to make it compatible with simulation results, since the SIDM simulations agree with NFW profile at large radii but are clearly different at small radii. These modifications include: $(i)$ a power change in the denominator of Eq.~(\ref{NFW}) to form a cored behavior at inner region, and $(ii)$ adding a new term in its numerator to reduce the size of cored region. Fig. (\ref{core}) shows how the profile changes without the second modification. This new term consists of a constant $k$ and a function $\Psi=\Psi (\alpha)$ which should be found. We should determine $\Psi$ such that as $\alpha$ goes from one to zero, $\Psi$ gives suitable numerical values that smoothly change the profile from the NFW to a cored shape. After calculating these values for many various inputs of $\alpha$, the function $\Psi (\alpha)$ is found by fitting it to the numerical result, whereby the answer is
\begin{equation} \label{profile_parameters}
\Psi=(1-\alpha ^{1-\sqrt{\alpha}}) \, [0.05 \, \exp(7.5\alpha)+10] \, .
\end{equation}
Fig. (\ref{Psi_fig}) illustrates how this function changes with $\alpha$. This relation has $RMSE=0.008$, which indicates a good accuracy for the fitted model (the statistical parameter of $RMSE$ is introduced in Appendix A). \\

Finally, $\alpha$ is the only parameter that represents a physical meaning and characterizes the density profile for each input of $\frac{\rm{M}}{\rm{M}_{\odot}}$ and $\frac{\sigma}{m}$. In fact, we tried to find the density profile just as a function of a single variable that represents the physical properties of SIDM halos. As mentioned before, the simulations reveal the dipendency of density profile (and thus $\alpha$) to cluster mass and cross section per unit mass of DM particles. As in our model $0 < \alpha < 1$, and given the simulations, $\frac{\sigma}{m}$ has a stronger effect than the cluster mass, we suggest the following relation as an ansatz,

\begin{equation} \label{alpha}
\alpha=1-\sigma_m (\frac{\rm{M}}{10^{15} \; \rm{M}_{\odot}})^A \, .
\end{equation}

We introduce the dimensionless cross-section per unit mass of $\sigma_m=\frac{\sigma/m}{1 \; \rm{cm}^2 g^{-1}}$ to have a dimensionless parameter of $\alpha$ (the value of $1 \; \rm{cm}^2 g^{-1}$ is chosen in the denominator since it is the upper limit of $\frac{\sigma}{m}$ in our study). Here, $A$ is a new constant that determines how much the change in slope of the profile depends on cluster mass at small radii, in comparison with the role of $\sigma_m$. A larger value of $A$ reveals a higher dependency on cluster mass. Having mentioned that, $A$ should have a negligible positive value to be in agreement with the simulation result. Similar to $\sigma_m$, the denominator in this term is considered to be $10^{15} \; \rm{M}_{\odot}$ to confine the whole term to changing between zero and one. \\

Clearly, $\alpha=1$ in Eq.~(\ref{density_profile}) results in NFW profile, while $\alpha=0$ describes a completely cored profile. In other words, we have designed the above ansatz in a way that it reduces to NFW profile if $\frac{\sigma}{m}$ is negligible, and shows a cored profile for the upper limit of $\frac{\sigma}{m}$ and the highest mass in the given range. \\

\begin{figure}[h]
 	\centering
 	\includegraphics [width=0.7 \linewidth] {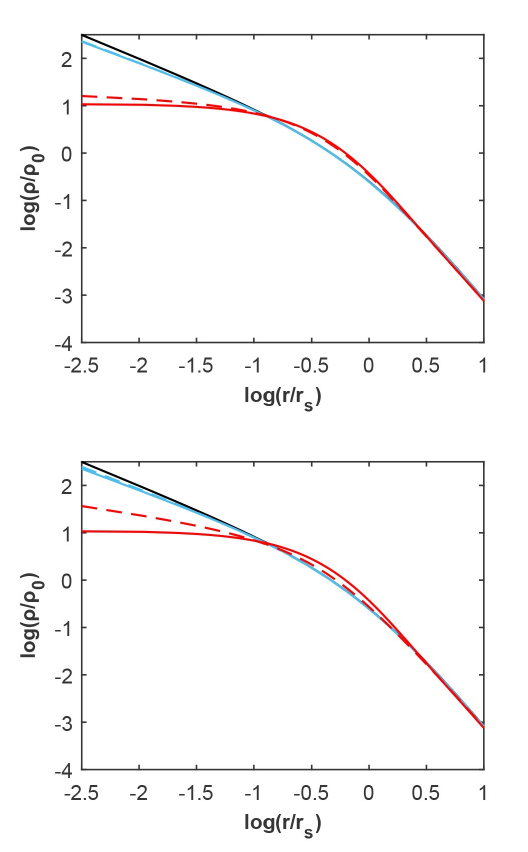}
 	\caption{Density profile of SIDM cluster sized halos, regarding the function of Eq.~(\ref{density_profile}) with $k=4$ and two choices of $A$ (top plot for $A=0.05$ and the bottom one for the case of $A=0.2$). Black color shows the NFW profile, while SIDM0.1 and SIDM1 are represented by blue and red curves, respectively. For both the cases, solid curves illustrate $\rm{M}=10^{15}\rm{M}_{\odot}$, while dashed curves stand for $\rm{M}=10^{14}\rm{M}_{\odot}$.}
 	\label{figure_1}
 \end{figure}

Fig. (\ref{figure_1}) illustrates Eq.~(\ref{density_profile}) for two values of $A=0.05$ and $A=0.2$ and also compares this function with NFW profile. In this figure (and following figures, too), blue curves show SIDM0.1 and red curves represent SIDM1. As can be seen, the suggested function for the density profile is in good agreement with the simulation of Ref. \cite{Robertson:2018anx} (see Appendix B for further detail). Both mass and $\frac{\sigma}{m}$ make a contribution to the changes in slope of density profile at the inner region, in the way that it is totally consistent with simulation results, and therefore, this function can be used for the density profile of SIDM galaxy clusters. A more exact value of $A$ should be obtained via iterating simulations for different cluster masses and through fitting the result to Eq.~(\ref{alpha}). Owing to the fact that we just have access to two windows of mass, we can just guess the value of this constant. We restrict our study to these two mentioned values of $A$ but note that its precise value can only be determined through performing simulation for a number of different masses. \\

\begin{figure}[h]
 	\centering
 	\includegraphics [width=0.7 \linewidth] {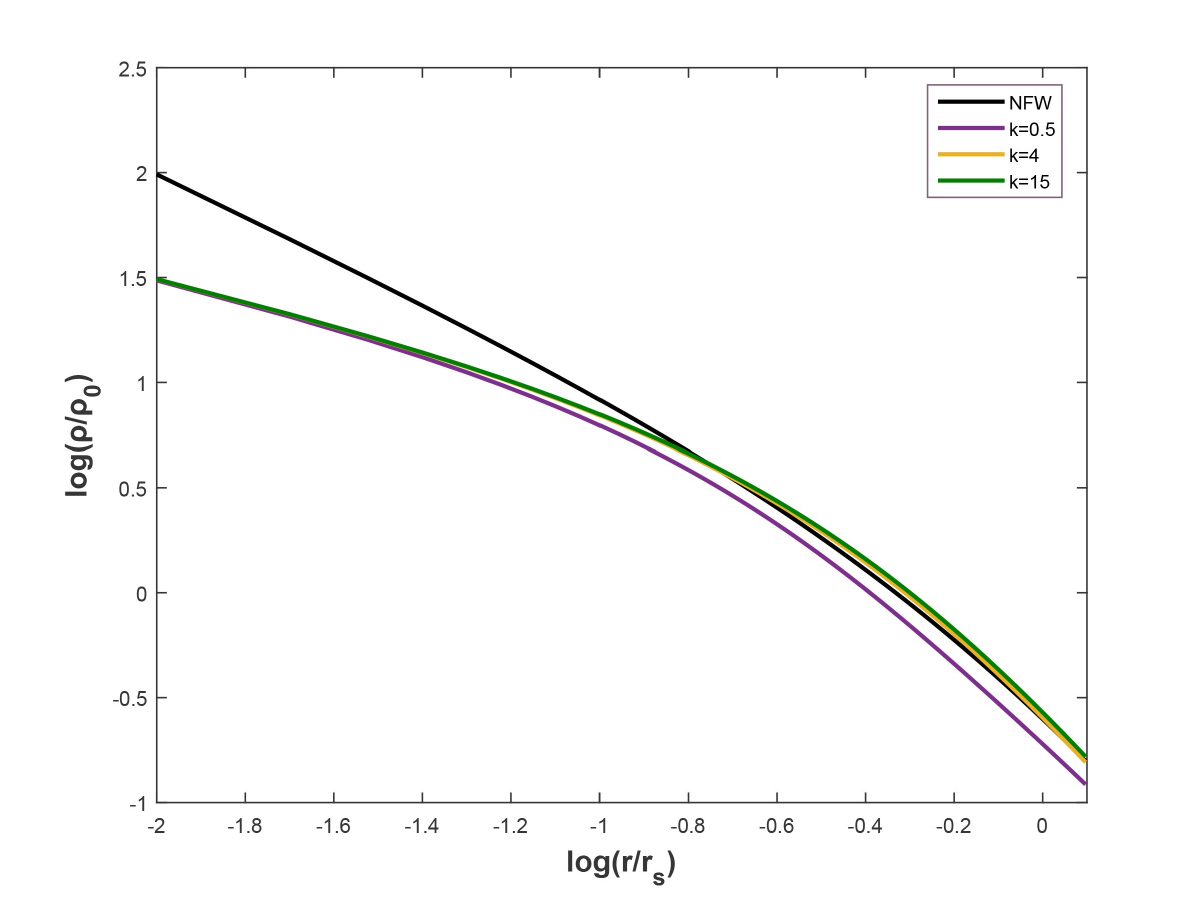}
 	\caption{Density profile of Eq.~(\ref{density_profile}) with $\alpha=0.5$ and three values of $k=0.5$ (purple line), $k=4$ (yellow line) and $k=15$ (green line). Black line shows the NFW profile.}
 	\label{figure_3}
 \end{figure}

In Fig. (\ref{figure_1}), we used $k=4$ for all curves. This preference is kept throughout this paper because it is an appropriate choice that leads the profiles to the NFW profile at large radii, completely similar to simulation results. To have a frame of reference, Fig. (\ref{figure_3}) reveals how different values of $k$ may change the behavior of the profile. Among three illustrated choices of $k=0.5, 4, 15$, the best choice that would fit the behavior of the function to the NFW profile is $k=4$. \\

Note that the simulations of Refs. \cite{Brinckmann:2017uve} and \cite{Robertson:2018anx} (and also our results) are related to the halos at $z\approx 0$. Since SIDM halos evolve with time, one of the main differences between CDM and SIDM halos is the fact that while the CDM density profile of halos remains stable after virialization, the SIDM density profile continuously changes. A time evolution type of simulation for these halos has been performed by Fischer et al. \cite{Fischer:2020uxh}. It shows that at the time of virialization, the SIDM density profile is consistent with the NFW profile (Fig. 5 of Ref. \cite{Fischer:2020uxh}). After this time, the density profile changes at small radii and the core is gradually formed until a particular time, which depends on the value of $\frac{\sigma}{m}$. After the core formation, self-interactions among DM particles result in energy loss inside the inner region of the halo and consequently, the core becomes smaller and denser (the gravothermal core-collapse). With regard to this behavior, our equations are restricted to low-redshift clusters which have not undergone the core-collapse yet. Another significant conclusion of this behavior is that self-interactions among DM particles do not affect virialization (density profile of SIDM halos overlaps with the NFW profile at the time of halo formation). Thus, the equations that are derived for CDM halos at virialization time, e.g. Eq.~(\ref{overdensity}), are still valid for the SIDM case. \\

It should also be mentioned that there are many other cored density profiles that have been proposed so far, such as Burkert \cite{Burkert:1995yz} and Hernquist \cite{Hernquist}. Nonetheless, we need a profile that slowly changes from a cuspy form to a cored shape. To this purpose, previous density profiles cannot be used as a suitable choice to describe galaxy clusters with SIDM, although they would be used for galaxies. \\

\subsection{Mass-Concentration Relation}

As mentioned before, the overdensity of a virialized halo in CDM model is near $\approx 200$. Therefore, the concentration parameter of a halo is defined with the relation of 

\begin{equation} \label{concentration}
c=\frac{r_{200}}{r_s} \approx \frac{r_{vir}}{r_s} \, ,
\end{equation}

as a function of viral radius and scale radius in the NFW profile. The relation between mass and concentration can be written in the form of

\begin{equation} \label{m_c_NFW}
c=a(\frac{\rm{M}_{200}}{10^{12} \; h^{-1}\rm{M}_{\odot}})^{b} \, ,
\end{equation}

where Hubble parameter is given by $H=100 \; h$ and the quantities of $a$ and $b$ vary in different studies. One of the most common forms of mass-concentration (M-c) relation in CDM model is given by $a=8.3$ and $b=-0.104$ according to Macciò et al. \cite{Maccio:2008pcd}. \\

Regarding the definition of $r_s$, scale radius is the distance from the center of the halo at which the logarithmic slope of the density profile equates to $-2$, meaning that $\rho(x) \propto x^{-2}$. We may use this definition to obtain a concentration parameter for the profile of Eq.~(\ref{density_profile}). Considering our density profile of $\rho(x)$, we can obtain $x_s$ (dimensionless radius where results in $\rho(x_s) \propto x^{-2}$) via calculating logarithmic derivative of density profile as

\begin{equation} \label{x_s}
\frac{d}{dlnx}(ln\rho(x))|_{x=x_s}=\frac{d}{dlnx}(ln(x^{-2}))|_{x=1}=-2 \, .
\end{equation}

As it was clarified at the end of the previous subsection, virialization is not affected by DM interactions in the SIDM model, and physical parameters, such as virial radius, are the same in SIDM and CDM halos at that time. Whereas, density profile in the SIDM model evolves with time, and consequently, $r_{200}$ gradually changes. The concentration parameter which is used in this study is $c=c_{vir}=\frac{r_{vir}}{r_s}$. Considering Eq.~(\ref{concentration}), it is possible to find the concentration parameter in our new profile as a function of the concentration of NFW profile ($c_{NFW}$) and $x_s$ of the previous equation via

\begin{equation} \label{c_new1}
\frac{c_{NFW}}{c}=\frac{r_s}{r_{s,NFW}}=x_s \, ,
\end{equation}

and hence

\begin{equation} \label{c_new}
c=\frac{c_{NFW}}{x_s} \, .
\end{equation}

\begin{figure}[h]
 	\centering
 	\includegraphics [width=0.7 \linewidth] {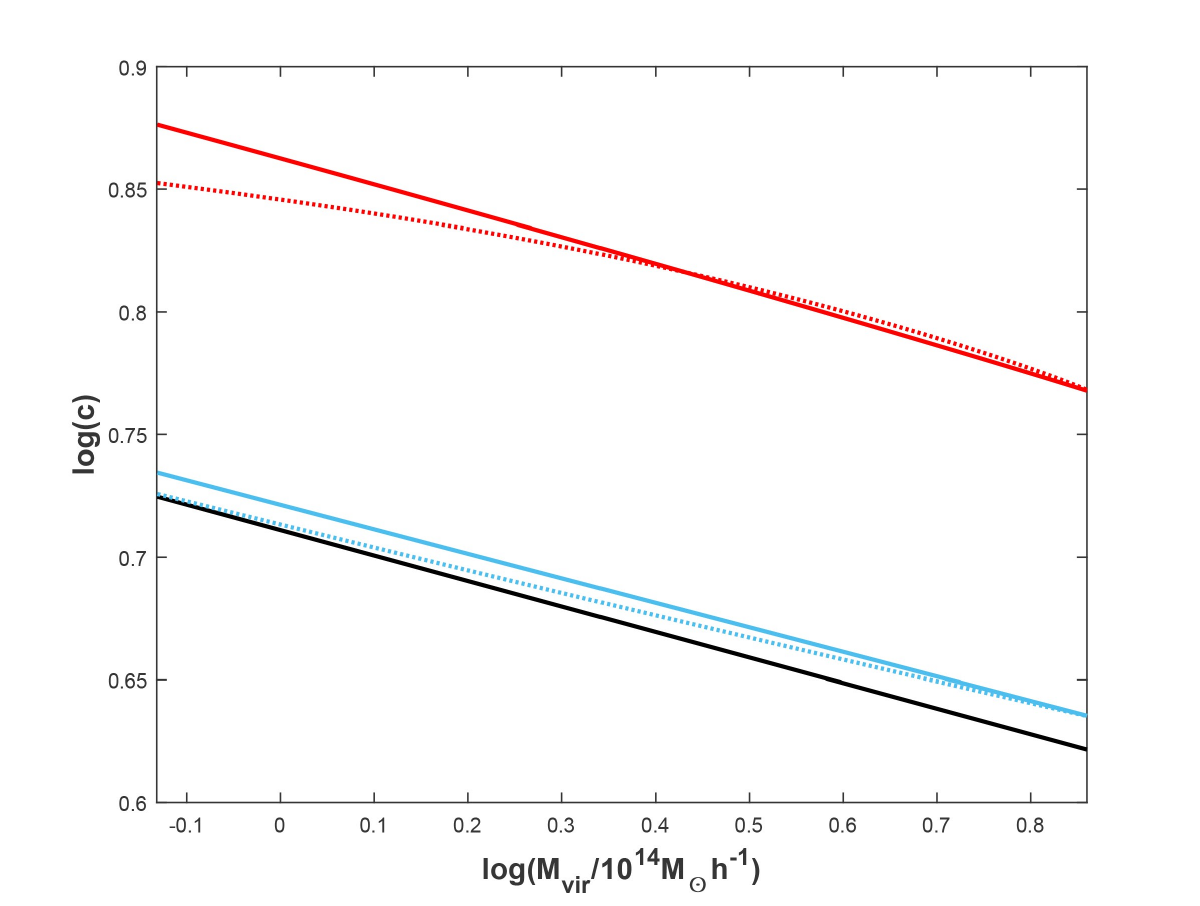}
 	\caption{Mass-concentration relation of the introduced density profile. Black line describes the M-c relation of Macciò et al. \cite{Maccio:2008pcd}, while blue and red lines stand for SIDM0.1 and SIDM1. Solid type of line denotes the case of $A=0.5$, and the other case with $A=0.2$ is represented by dotted lines.}
 	\label{figure_4}
 \end{figure}

This method helps to calculate the concentration parameter for the new density profile. The result is provided in Fig. (\ref{figure_4}) for M-c relation and it reveals that SIDM leads to an upward shift in the M-c plot. The value of $\frac{\sigma}{m}$ determines how remarkable this shift would be. Furthermore, dependency of $\alpha$ on mass (which is expressed by $A$) alters the slope of the M-c relation. In order to obtain the relevant M-c relation, we fit these plots to Eq.~(\ref{m_c_NFW}). The results are summarized in Table~\ref{table_0}. \\

\begin{table}[h]
\caption{Numerical values obtained by fitting the plots of Fig. (\ref{figure_4}) to Eq.~(\ref{m_c_NFW}), in order to find the constants of $a$ and $b$ in this equation.}
\begin{tabular}{ccccl}
\cline{3-4}
                                              & \multicolumn{1}{c|}{}        & \multicolumn{1}{c|}{$a$}      & \multicolumn{1}{c|}{$b$}      &  \\ \cline{1-4}
\multicolumn{1}{|c|}{\multirow{2}{*}{A=0.05}} & \multicolumn{1}{c|}{SIDM0.1} & \multicolumn{1}{c|}{$8.34$} & \multicolumn{1}{c|}{$-0.100$}  &  \\ \cline{2-4}
\multicolumn{1}{|c|}{}                        & \multicolumn{1}{c|}{SIDM1}   & \multicolumn{1}{c|}{$12.09$}  & \multicolumn{1}{c|}{$-0.110$}  &  \\ \cline{1-4}
\multicolumn{1}{|c|}{\multirow{2}{*}{A=0.2}}  & \multicolumn{1}{c|}{SIDM0.1} & \multicolumn{1}{c|}{$7.85$} & \multicolumn{1}{c|}{$-0.091$}  &  \\ \cline{2-4}
\multicolumn{1}{|c|}{}                        & \multicolumn{1}{c|}{SIDM1}   & \multicolumn{1}{c|}{$10.63$}   & \multicolumn{1}{c|}{$-0.088$} &  \\ \cline{1-4}
\multicolumn{1}{l}{}                          & \multicolumn{1}{l}{}         & \multicolumn{1}{l}{}        & \multicolumn{1}{l}{}        & 
\end{tabular}
\label{table_0}
\end{table}

Note that this upward change has also been obtained in several other studies for SIDM1 \cite{Bondarenko:2017rfu} and also for a velocity-dependent cross-section model of SIDM \cite{Sagunski:2020spe}.

\subsection{Velocity Dispersion}

SIDM simulation \cite{Robertson:2018anx} also provides a velocity dispersion profile of halos. In terms of a qualitative discussion, velocity dispersion in a CDM halo increases with the radius, hitting its peak at $r_{max}\approx 2.1 r_s$, and then declines until a very low value at high radii. Inside SIDM halo, DM collisions give rise to heat transfer from the region with the highest velocity dispersion ($r_{max}$) to the inner region of halo, resulting in an increase in kinetic energy (and thus, velocity dispersion) in the central part of the cluster. This phenomenon affects velocity dispersion profile and as can be seen in the third set of plots in Fig. (2) of Ref. \cite{Robertson:2018anx}, velocity dispersion approximately remains at its maximum until $r_{max}$ of CDM and then falls, similar to CDM case. \\

In order to quantify such a behavior, one can solve the Jeans equation \cite{Sokolenko:2018noz} for 3-dimensional velocity dispersion ($\sigma_v$), which is written as 

\begin{equation} \label{Jeans}
\frac{d}{dr}\left(\frac{\sigma_v^2}{3} \frac{r^2}{\rho} \frac{d\rho}{dr}\right)=-4 \pi G r^2 \rho \, .
\end{equation}

To solve this first-order differential equation (using the density profile of Eq.~(\ref{density_profile})), a boundary condition is required. At large radii ($r\to \infty$), velocity dispersion goes to zero. Nonetheless, it would not be used as an appropriate boundary condition, owing to the fact that all halos (with any value of $\alpha$) should satisfy this condition regarding simulation results. We use some approximations to find a suitable initial condition to solve this differential equation. \\

For SIDM0.1, we can infer that there is only a negligible difference from CDM outcome. We neglect this difference and consider that SIDM0.1 velocity dispersion profile is similar to that of CDM. The value of 1-dimensional velocity dispersion ($\sigma_{1D}$) at virial radius can be obtained for CDM halos via \cite{Evrard:2007py}

\begin{equation} \label{velocity_dispersion_CDM}
\frac{\sigma_{1D}}{\rm{km}/\rm{s}}=1082.9 \; \left(\frac{\rm{M}_{200}}{10^{15} \; h^{-1}\rm{M}_{\odot}}\right)^{0.3361} \, .
\end{equation}

Assuming that velocity dispersion profile is a symmetric function relative to the vertical axis at $r_{max}$, we can claim that the obtained value of Eq.~(\ref{velocity_dispersion_CDM}) is also the value of $\sigma_{1D}$ at $r_{init}$, where we choose as the initial condition to solve the Jeans equation. The radius of $r_{init}$ is then determined by

\begin{equation} \label{r_init_1}
\log(r_{max})-\log(r_{init})=\log(r_{vir})-\log(r_{max}) \, ,
\end{equation}

which reduces to 

\begin{equation} \label{r_init_2}
\log(x_{init})=2 \; \log(x_{max})-\log(c) \, ,
\end{equation}

where Eq.~(\ref{concentration}) is used. Finally, putting the rough value of $x_{max} \approx 2.1$ in above equation, $x_{init}$ is obtained via 

\begin{equation} \label{r_init}
x_{init}=\frac{2.1^2}{c} \, .
\end{equation}

\begin{figure}[h]
 	\centering
 	\includegraphics [width=0.7 \linewidth] {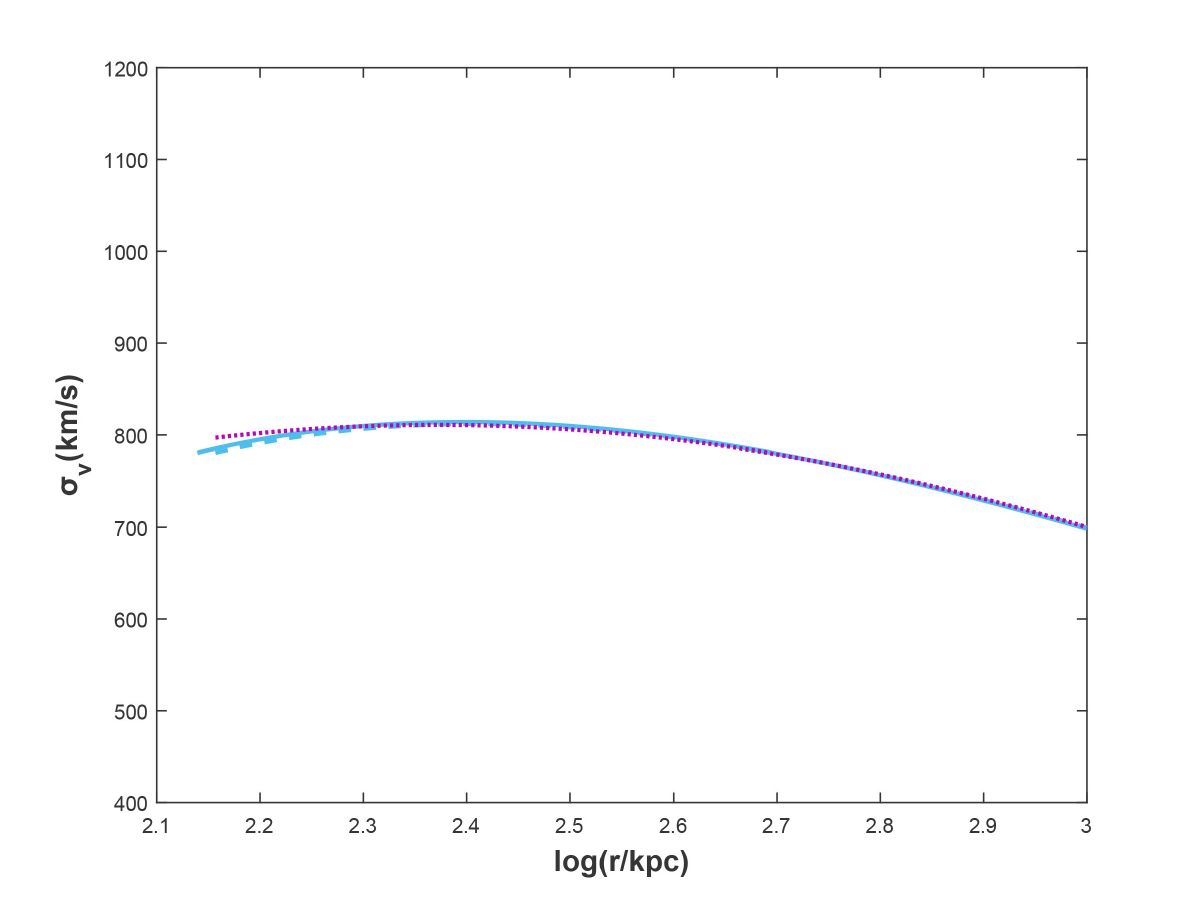}
 	\caption{The 3-dimensional velocity dispersion profile in a SIDM0.1 halo as a function of radius. Solid blue curve represents the solution of the Jeans equation for $A=0.05$. This solution for $A=0.2$ is provided by dashed blue line, although it is not easily distinguishable from the case of $A=0.05$ in this plot. Fitted Gaussian curve is illustrated by dotted magenta line, which is in a very good agreement with the solution.}
 	\label{figure_5}
 \end{figure}

This procedure leads to finding a velocity dispersion profile, although $x_{init}$ is quite large whereby we lose the substantial area in which the SIDM effect differs from CDM. As a solution to this problem, one can fit a Gaussian function of 

\begin{equation} \label{Gaussian}
\frac{\sigma_v}{\rm{km}/\rm{s}}=a e^{-(\frac{t-b}{c})^2} \, ,
\end{equation}

to the solution of the Jens equation, where $t=\log(r/\rm{kpc})$ and the constants of $a$, $b$ and $c$ vary for profiles with different values of $\alpha$. Gaussian function can be a very good fit to the solution of the Jeans equation in SIDM0.1. Such an assertion could be proved in Fig. (\ref{figure_5}), which illustrates both solution of the Jeans equation and its fitted Gaussian function for SIDM0.1 in a halo with $\frac{\rm{M}}{\rm{M}_{\odot}}=10^{14}$. It is clear that the fitted Gaussian function (dotted curve) with $a=810.9$, $b=2.372$ and $c=1.639$ is consistent with the solution of Eq.~(\ref{Jeans}) with regard to the mentioned initial condition, which is represented by blue curve. Note that we used $\sigma_v \approx \sqrt{3} \sigma_{1D}$ and neglected velocity anisotropy, even though there could emerge different velocity anisotropy values at different radii of the halo \cite{Brinckmann:2017uve}. \\

Another result of Fig. (\ref{figure_5}) is a very negligible difference between two cases of $A=0.05$ and $A=0.2$, thereby these two cases overlap in a plot. \\

\begin{figure}[h]
 	\centering
 	\includegraphics [width=0.7 \linewidth] {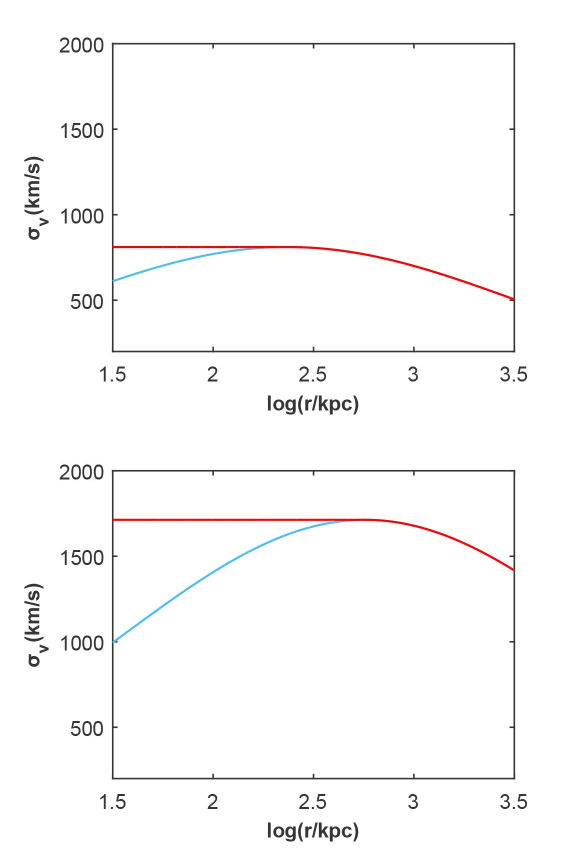}
 	\caption{The 3-dimensional velocity dispersion as a function of radius in a halo with the mass of $\frac{\rm{M}}{\rm{M}_{\odot}}=10^{14}$ (top plot) and $\frac{\rm{M}}{\rm{M}_{\odot}}=10^{15}$ (bottom plot). SIDM0.1 is illustrated with blue curve and red curve represents SIDM1.}
 	\label{figure_6}
 \end{figure}

For SIDM1, we assume that at $r<r_{max}$, velocity dispersion remains unchanged, having its value at $r_{max}$ and then, overlaps with its relevant Gaussian function in SIDM0.1. Fig. (\ref{figure_6}) shows 3-dimensional velocity dispersion profiles of SIDM0.1 (blue) and SIDM1 (red) for $\frac{\rm{M}}{\rm{M}_{\odot}}=10^{14}$ and $\frac{\rm{M}}{\rm{M}_{\odot}}=10^{15}$, respectively, in which the explained method is used. It is clear that these plots are in good agreement with simulation and consequently, this procedure can be used in Section III to derive scaling relations. 

\section{Scaling Relations}
%

In order to completely characterize a cluster with observational data, scaling relations would be needed. Specifically, cluster masses are difficult to be directly measured, and therefore, mass-temperature (M-T) and mass-velocity dispersion (M-$\sigma_v$) relations make a notable contribution to determining fundamental features of galaxy clusters. In this Section, we investigate these two important relations in the SIDM model. \\

\subsection{Mass-Velocity Dispersion Relation}

Once we have the velocity dispersion profile, we can immediately find a relation between mean velocity dispersion and mass of the clusters. Considering $\bar{\sigma_v}$ to be the 3-dimensional mass-weighted mean velocity dispersion inside the radius of $r_{vir}$, this parameter can be written as

\begin{equation} \label{mean}
\bar{\sigma_v} = \frac{\int_0^c 4 \pi x^2 \rho (x) \sigma_v dx}{\int_0^c 4 \pi x^2 \rho (x) dx} \, .
\end{equation} 

\begin{figure}[h]
 	\centering
 	\includegraphics [width=0.7 \linewidth] {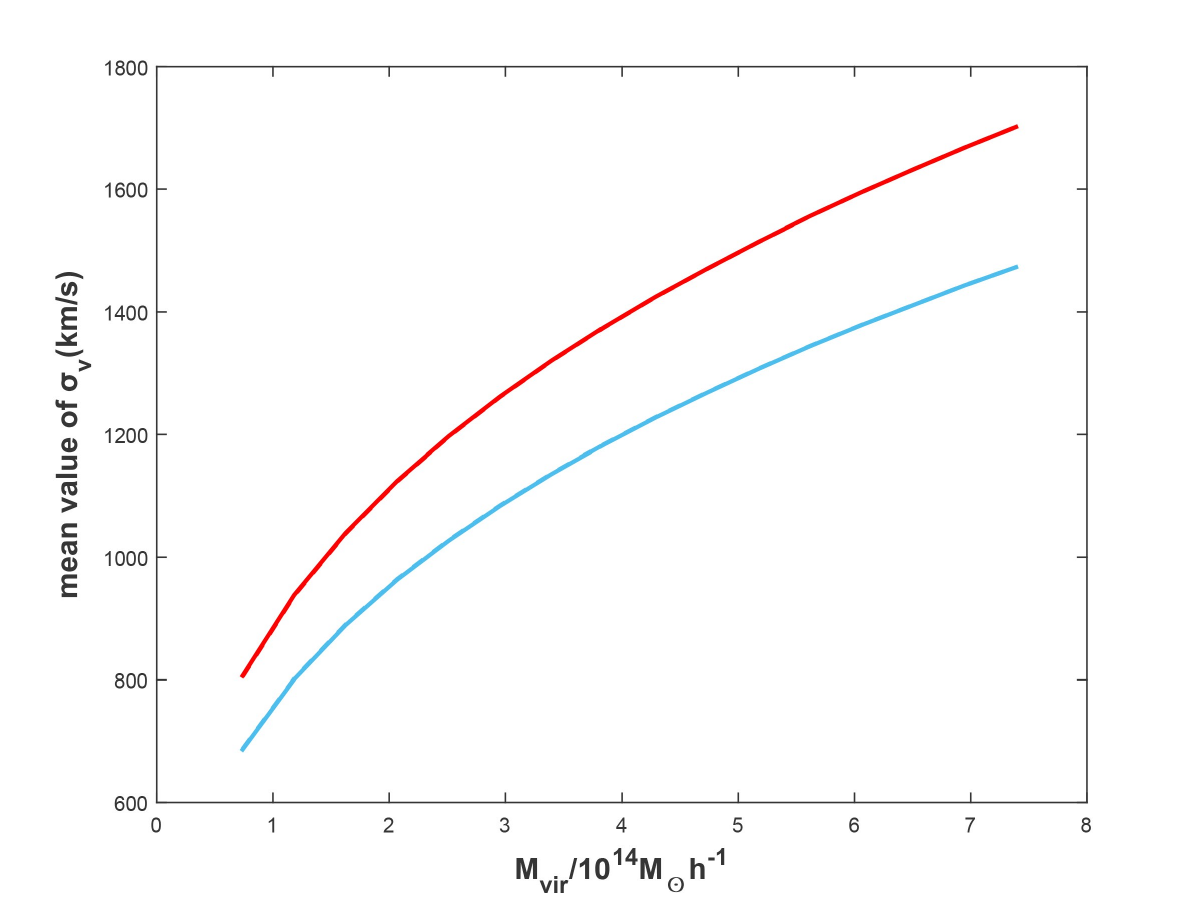}
 	\caption{Mass-velocity dispersion relation for SIDM0.1 (blue) and SIDM1 (red) models. Solid curves denote $A=0.05$ and dashed curves stand for $A=0.2$. However, the difference between these two cases is very negligible and approximately cannot be seen.}
 	\label{figure_8}
 \end{figure}

Using this equation, it is possible to find $\bar{\sigma_v}$ as a function of cluster mass. As can be seen in Fig. (\ref{figure_8}), a higher value of cross-section leads to a larger $\bar{\sigma_v}$ for the same mass. This is not surprising, as could already be inferred from Fig. (\ref{figure_6}). \\

The curves in Fig. (\ref{figure_8}) can also be fitted to suitable functions to have a quantitative mass-velocity dispersion relation. To this purpose, we find the relation of

\begin{equation} \label{M_C_0.1}
\frac{\bar{\sigma_v}}{\rm{km}/\rm{s}} = 1628 \; (\frac{\rm{M}_{vir}}{10^{15} \; h^{-1}\rm{M}_{\odot}})^{0.3328} \, 
\end{equation}

for SIDM0.1, and similarly the relation of

\begin{equation} \label{M_C_1}
\frac{\bar{\sigma_v}}{\rm{km}/\rm{s}} = 1876 \; (\frac{\rm{M}_{vir}}{10^{15} \; h^{-1}\rm{M}_{\odot}})^{0.3249} \, 
\end{equation}

is obtained for SIDM1. For the 16 numerical values of $10^{14}<\frac{\rm{M_{vir}}}{\rm{M}_{\odot}}<10^{15}$ that were used to fit the last two relations, we have $RMSE=1.065$ and $RMSE=0.7518$, respectively, which are negligible compared to the order of $\bar{\sigma_v}$. According to Fig. (\ref{figure_8}), plots of $A=0.05$ and $A=0.2$ overlap with each other and as a result, the above relations can be used for both these cases of $A$.

\subsection{Mass-Temperature Relation}

The mass-temperature relation plays a crucial role in determining cluster masses. Mass of a cluster can be found via several methods, including gravitational lensing. However, the M-T relation can pave the way for a more straightforward way of mass measurement. In our previous work \cite{Naseri:2020uvn}, we followed the method provided by Afshordi \& Cen \cite{Afshordi:2001ze} and derived the M-T relation with regard to the interacting dark sector model in which a possible interaction is assumed between dark matter and dark energy. Here, we pursue that method of obtaining M-T relation and then investigate its modifications in the SIDM model. \\

This procedure starts with the virial theorem and combines it with the conservation of energy to find a relation between mass and temperature. While in an interacting dark sector model, the virial ratio changes and results in the main differences \cite{Naseri:2020uvn}, it would not be remodeled in SIDM. In fact, the possible interaction between DM particles has nothing to do with the virial theorem, as a variety of interactions among baryons might not lead to a different virial condition (for a review, you may follow Section II in Ref. \cite{Naseri:2020uvn}). In a SIDM halo, none of the fundamental assumptions and principles would change and thus, the M-T relation keeps its form in the $\Lambda$CDM model, that is

\begin{equation}\label{MTQ}
k_BT = (6.62 \; \rm{keV}) \tilde{Q} \left(\frac{\rm{M}}{10^{15} \;h^{-1} \rm{M}_{\odot}}\right)^{2/3}\, ,
\end{equation}
 where $k_B$ is the Boltzmann constant and $\tilde{Q}$ is a dimensionless factor. This factor contains every detail of SIDM inside itself and is written as

\begin{equation}\label{Q}
\tilde{Q} = (\frac{\tilde{\beta}_{spec}}{0.9})^{-1}\left(\frac{1+\nu}{1-\nu}\right)y\, ,
\end{equation}

with

\begin{equation}\label{y_relation}
 y(c,f(x)) = \frac{\Delta^{1/3}(1-\nu)c \int_0^c f(x) g(x) x dx}{3\pi^{2/3} g^2(c)}\, .
\end{equation}

Here, $\Delta$ is the overdensity of the virialized halo, and $\nu$ is a coefficient constant that has emerged since the surface pressure term on the boundary of the cluster ($P_{ext}$) is taken into account in virial condition. This constant relates $P_{ext}$ to the potential energy of virialized sphere ($U$) with volume of $V$ via \cite{Afshordi:2001ze} 

\begin{equation}\label{nu}
\nu(c,f(x)) = -\frac{3P_{ext}V}{U}= \frac{c^3\int_c^{\infty} f(x)g(x)x^{-2}dx}{\int_0^c f(x) g(x) x dx}\, .
\end{equation}

It is self-evident that Eqs.~(\ref{y_relation}) and (\ref{nu}) are highly dependent on the modified density profile ($f(x)$) and concentration parameter of SIDM halo. Not only does the effect of density profile directly appears in the above equations, but it also affects another function $g(x)$, which has the definition of

\begin{equation}\label{g}
g(x) = \int_0^x f(x) x^2 dx \, .
\end{equation}

Another significant contributor to the M-T relation is $\tilde{\beta}_{spec}$, which has the form of

\begin{equation}\label{tilde_Beta}
\tilde{\beta}_{spec} = \beta_{spec} [1 + (f \beta_{spec}^{-1}-1)\frac{\Omega_b}{\Omega_b+\Omega_{DM}}]\, ,
\end{equation}

where $f$ is  the fraction of baryonic matter in hot gas and is considered to be near one \cite{Afshordi:2001ze}, $\Omega_b$ and $\Omega_{DM}$ are the relevant density parameters of baryonic matter and DM, and the spectroscopic beta parameter is defined as

\begin{equation}\label{Beta_spec}
\beta_{spec} =\frac{\sigma_{1D}^2}{(k_BT/\mu m_p)} \, .
\end{equation}

Here, $\mu=0.59$ is the mean molecular weight and $m_p$ denotes the proton mass. The impact of velocity dispersion emerges in Eq.~(\ref{Beta_spec}) and we should use Eqs.~(\ref{M_C_0.1}) and (\ref{M_C_1}) in that for SIDM0.1 and SIDM1, respectively. Given these equations, it is plausible to have a relation between mass and temperature of a cluster. As a result, we obtain corresponding temperatures for given masses between $10^{14}<\frac{\rm{M}}{\rm{M}_{\odot}}<10^{15}$ with the aid of above equations. Fitting the relation of 

\begin{equation}\label{M_T_fitting}
\frac{\rm{M}}{10^{14} \; h^{-1} \rm{M}_{\odot}}=p \; \left(\frac{k_BT}{\rm{keV}}\right)^q \, ,
\end{equation}

to the calculated T values for input of M values gives constants of $p$ and $q$ and consequently, we can find the M-T relation for SIDM model. The result is summarized in Table~\ref{table_1} and it can be seen that for SIDM1, the power-law between mass and temperature faces a significant change, both in the power index and the normalization factor. \\

\begin{figure}[h]
 	\centering
 	\includegraphics [width=0.7 \linewidth] {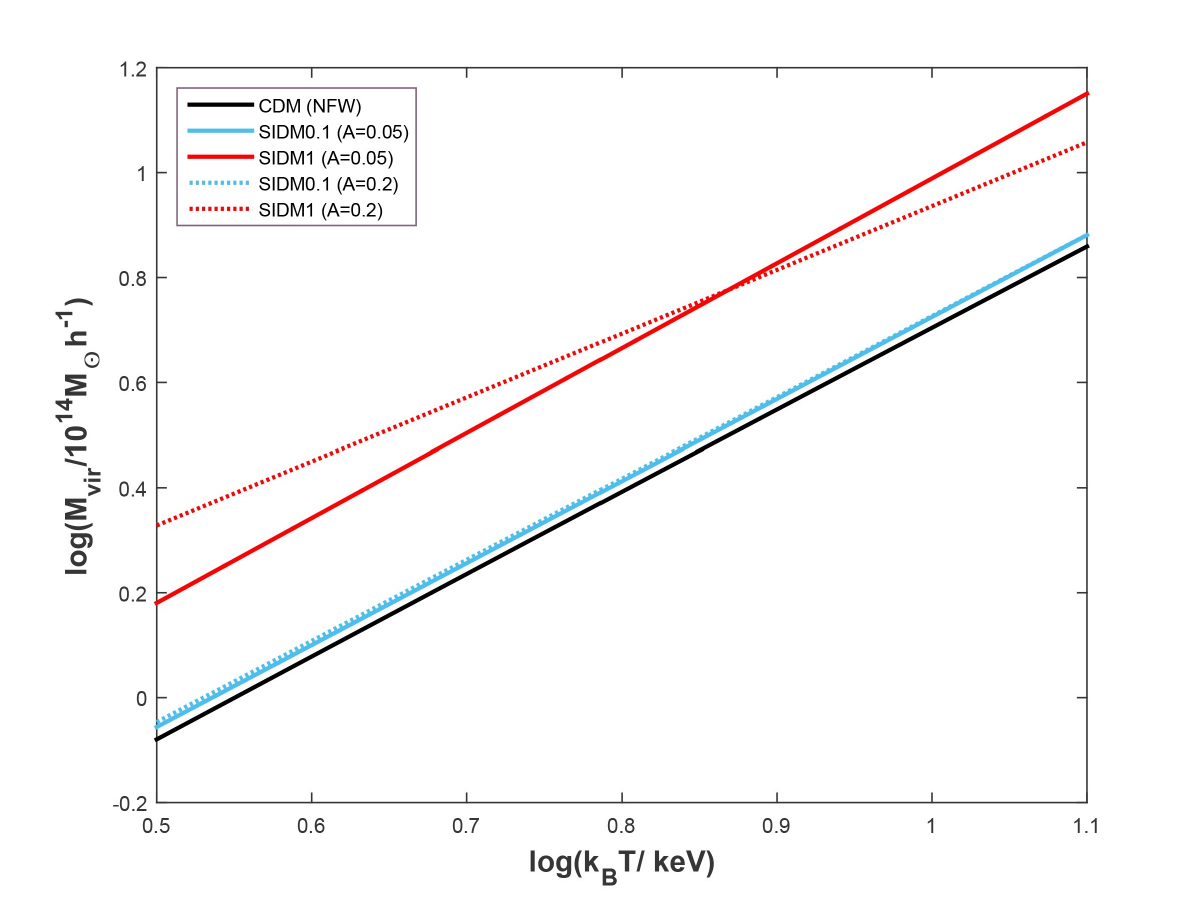}
 	\caption{Mass-temperature relation for SIDM0.1 (blue line) and SIDM1 (red line) models. Solid and dotted lines denote $A=0.05$ and $A=0.2$, respectively. Moreover, the result of the NFW profile in provided by black line.}
 	\label{figure_9}
 \end{figure}

Using this method, the modified M-T relation is illustrated in Fig. (\ref{figure_9}). In this figure, SIDM0.1 is very close to the CDM result in the NFW profile. However, it is clear that the obtained mass for each given temperature increases in SIDM1 halos. It is an important conclusion, as the cluster masses are usually determined according to x-ray temperature. Although the difference between SIDM1 and CDM results is only around 0.2 order of magnitude, which is normal and exists among various observational datasets with diverse methods of mass measurement \cite{Naseri:2020uvn}, clusters should be more massive than what has been considered in the CDM model if DM particles extensively interact with each other. \\ 

In addition, high dependency of the density profile on mass (or higher $A$, equivalently) completely changes the behavior of M-T relation. Perhaps the modified M-c relation plays the most profound role in this change, due to its large difference from the other cases (take a look at Fig. (\ref{figure_4})). Strange behavior of M-c and M-T relations for SIDM1 in $A=0.2$ raises doubts about the high dependency of density profile on cluster mass. Overall, $A=0.05$ gives rise to more familiar results in these two equations, although it is not different than $A=0.2$ in mass-velocity dispersion relation. 

\begin{table}[h]
\caption{The numerical values of constants in Eq.~(\ref{M_T_fitting}) that are obtained via fitting this equation to calculated temperatures for 16 given masses in the range of $10^{14}<\frac{\rm{M}}{\rm{M}_{\odot}}<10^{15}$. The value of $RMSE$ is also reported.}
\begin{tabular}{cc|c|c|c|}
\cline{3-5}
                                              &         & $p$     & $q$     & $RMSE$ \\ \hline
\multicolumn{1}{|c|}{\multirow{2}{*}{$A=0.05$}} & SIDM0.1 & 0.148 & 1.554 & 0.0013    \\ \cline{2-5} 
\multicolumn{1}{|c|}{}                        & SIDM1   & 0.222 & 1.649 & 0.0041    \\ \hline
\multicolumn{1}{|c|}{\multirow{2}{*}{$A=0.2$}}  & SIDM0.1 & 0.154 & 1.539 & 0.0013    \\ \cline{2-5} 
\multicolumn{1}{|c|}{}                        & SIDM1   & 0.480 & 1.266 & 0.0078    \\ \hline
\end{tabular}
\label{table_1}
\end{table}

\section{Super Interacting Dark Sector}
%

In the previous Section, it was concluded that for SIDM1, corresponding mass to every input temperature is higher than that of SIDM0.1 and CDM, with difference of around 0.2 order of magnitude. It does not prove a large discrepancy, as various observational methods of mass measurement usually have differences in this range compared with each other. Nevertheless, in this Section, we show that the combination of a particular cosmological model of DE with the self-interacting DM can completely cancel the effect of such an upward shift in M-T relation. \\

The interacting dark sector is a DE-DM model in which a possible interaction is assumed between DE and DM. According to this model, DM particle mass is determined with regard to its interaction with a scalar field with the energy density of DE; similar to the attribution of quark and lepton masses to their interaction with the Higgs field \cite{Farrar:2003uw}. For the cosmology in which there is such an interaction between DM and DE, the mass of DM particles is directly determined according to the rate of energy transfer between DM and DE. Therefore, the value of $\frac{\sigma}{m}$ is affected by the interaction between DE and DM. \\

Given the assumptions of interacting dark sector model, cosmological equations, including the energy balance and the Raychadhuri field equations, change to the forms of 

\begin{eqnarray}\label{equations}
& \dot{\rho}_b = - 3H \rho_b \, ,  \nonumber \\ \nonumber \\
& \dot{\rho}_{DM} = - 3H \rho_{DM} + Q \, ,  \nonumber \\ \nonumber \\
& \dot{\rho}_{DE} = - 3(1+w_{DE}) H \rho_{DE} - Q \, ,  \nonumber \\ \nonumber \\
& \dot{H} = - 4\pi G \left[ \rho_b + \rho_{DM} + (1+w_{DE}) \rho_{DE} \right] \, .
\end{eqnarray}

Here $\rho_{DE}$, $\rho_{DM}$, and  $\rho_b$ denote energy density of DE, DM and baryonic matter, respectively, and $\omega_{DE}$ is constant of the equation of state for DE. Furthermore, $Q$ is the new term that describes the interaction between DE and DM. In other words, $Q$ indicates the rate of energy transfer between these two components; while a positive $Q$ shows the energy transfer from DE to DM, a negative $Q$ reveals this transfer from DM to DE. A number of functions have been proposed and investigated for $Q$ so far. One simple and useful choice is considering $Q$ to be proportional to the Hubble parameter. For this model, some of the studied functions are \cite{CalderaCabral:2008bx,vonMarttens:2018iav}

\begin{eqnarray}\label{Q_Models}
\mbox{ Model I} : Q = 3H (\alpha_c \rho_c + \alpha_x \rho_x) \, ,  \nonumber \\ \nonumber \\
\mbox { Model II} : Q = 3H \xi_1 \frac{\rho_c\rho_x}{\rho_c+\rho_x} \, ,  \nonumber \\ \nonumber \\
\mbox{ Model III} : Q = 3H \xi_2 \frac{\rho_x^2}{\rho_c+\rho_x} \, ,  \nonumber \\ \nonumber \\
\mbox{ Model IV} : Q = 3H \xi_3 \frac{\rho_c^2}{\rho_c+\rho_x} \, .
\end{eqnarray}

The parameters of $\xi_1$, $\xi_2$, $\xi_3$, $\alpha_{DE}$ and $\alpha_{DM}$ are the so-called interacting constants that describe each model, exclusively. \\

Considering the interacting dark sector model, the virial theorem turns into a different form. Most significantly, the virial ratio ($\lambda$) between kinetic energy $K$ and potential energy $U$ is not $\frac{1}{2}$ anymore. Writing the virial condition as 

\begin{equation} \label{virial}
K=-\lambda_i U \, ,
\end{equation}

the modified virial ratio becomes \cite{Naseri:2020uvn}

\begin{eqnarray}\label{VirialModels}
\mbox{ Model I} : \lambda_{I}=\frac{1-6\alpha_{DM}}{2+3\alpha_{DM}+3\alpha_{DE}/R}\, ,  \nonumber \\ \nonumber \\
\mbox{ Model II} : \lambda_{II}=\frac{1-\frac{6\xi_1}{R+1}}{2+\frac{3\xi_1}{R+1}}\, ,  \nonumber \\ \nonumber \\
\mbox{ Model III} : \lambda_{III}=\frac{1}{2+\frac{3\xi_2}{R(R+1)}}\, ,  \nonumber \\ \nonumber \\
\mbox{ Model IV} : \lambda_{IV}=\frac{1-\frac{6R\xi_3}{R+1}}{2+\frac{3R\xi_3}{R+1}}\, ,
\end{eqnarray}

where $i=I, II, III, IV$ and $R=\frac{\rho_{DM}}{\rho_{DE}}$. With this modified virial theorem, the coefficient factor of Eq.~(\ref{MTQ}) changes to 

\begin{equation}\label{modified_Q}
\tilde{Q} = \left(\frac{\tilde{\beta}_{spec}}{0.9}\right)^{-1}\left(\frac{2\lambda_i+\nu}{2-2\lambda_i-\nu}\right) y\, ,
\end{equation}

with

\begin{equation}\label{y_relation_two}
 y(c,f(x)) = \frac{\Delta^{1/3}(2-2\lambda_i-\nu)c \int_0^c f(x) g(x) x dx}{3\pi^{2/3} g^2(c)}\, . 
\end{equation}

Having these relations in mind, we now introduce a hybrid model which we call super interacting dark sector, or SIDS, in order to probe the variation of M-T relation in a hybrid and improved model. This model is founded on two fundamental assumptions: ($i$) DM particles have non-negligible cross-section per unit mass of $\frac{\sigma}{m}$ whereby they can interact with themselves (self-interacting DM) and ($ii$) DM particles interact with DE and their mass stems from this interaction (interacting dark sector). According to the second assumption, interacting constants ($\xi_1$, $\xi_2$, $\xi_3$, $\alpha_{DE}$ and $\alpha_{DM}$) are functions of DM mass and thereby, functions of $\frac{\sigma}{m}$. \\

SIDS is a simple combination of two different models of DE and DM. As interacting dark sector can solve many cosmological problems, e.g. the 'coincidence problem' \cite{CalderaCabral:2008bx}, and SIDM can provide solutions to some other astrophysical problems, such as core-cusp and missing satellite problems, SIDS could even be more effective and justify both groups of problems. There is no contradiction between the two basic models in theory and they can both be valid at the same time. In addition, in the interacting dark sector model, a Yukawa coupling of DE field to the DM particles is assumed as the basic idea \cite{Farrar:2003uw}, while one of the proposed types of interaction between DM particles in the SIDM model is Yukawa interaction \cite{Loeb:2010gj}. \\

\begin{figure}[h]
 	\centering
 	\includegraphics [width=0.7 \linewidth] {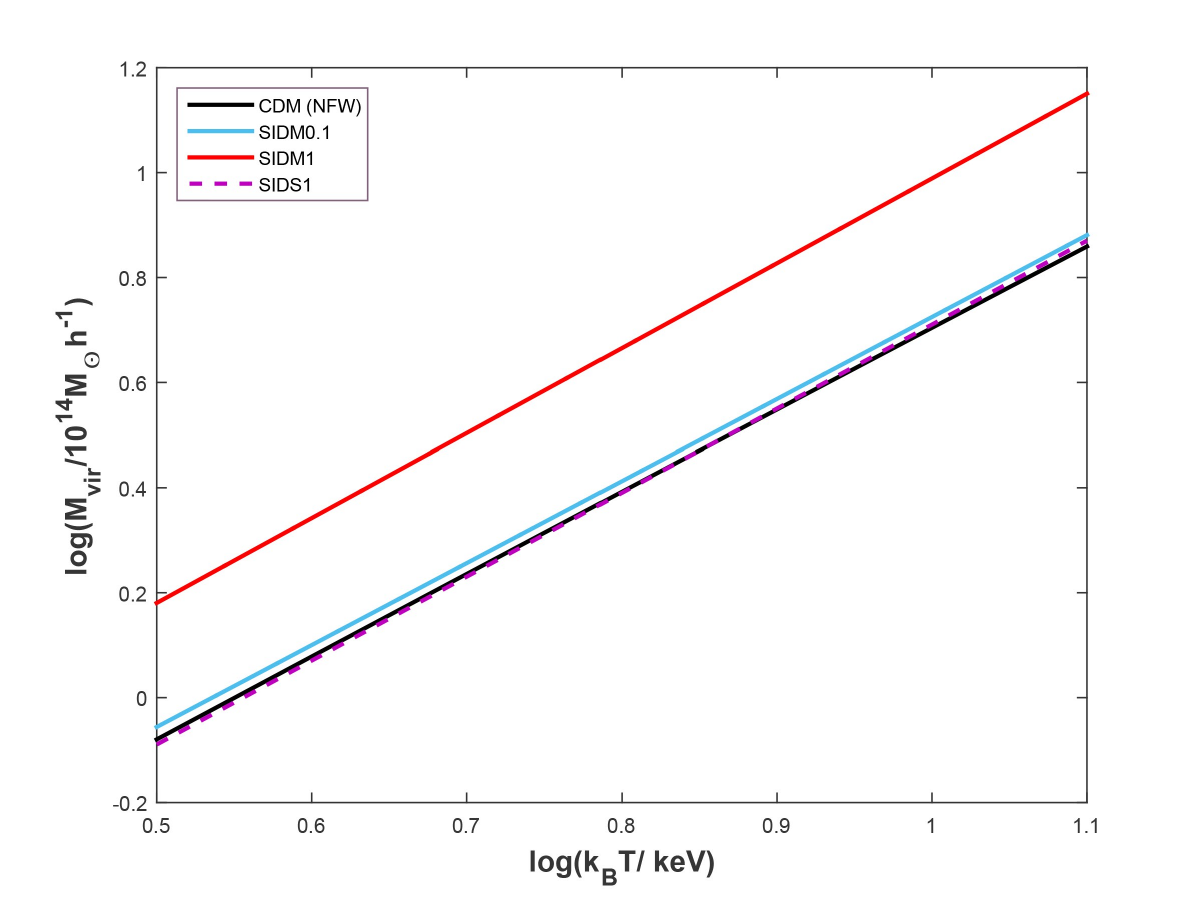}
 	\caption{Mass-temperation relation for SIDS in comparison with SIDM. Similar to the previous figure, SIDM1 and SIDM0.1 are shown with red and blue lines, respectively. The dashed magenta line illustrates SIDS model with $\frac{\sigma}{m}=1 \; \rm{cm}^2 g^{-1}$ and $\lambda =0.55$. All the lines are obtained for $A=0.05$.}
 	\label{figure_10}
 \end{figure}

To investigate M-T relation for the SIDS model, Eq.~(\ref{MTQ}) should be considered, again with density profile, concentration, and the velocity dispersion of SIDM, but also with adding the new term of modified virial ratio, i.e. Eqs.~(\ref{modified_Q}) and (\ref{y_relation_two}). Fig. (\ref{figure_10}) reveals how SIDS with $\lambda =0.55$ for $\frac{\sigma}{m}=1 \; \rm{cm}^2 g^{-1}$ omits the upward shift of the M-T plot. \\

For the given virial ratio of $\lambda =0.55$, it is possible to obtain interacting constants of Eq.~(\ref{VirialModels}) for specific models of Eq.~(\ref{Q_Models}). In addition, it was mentioned that the constants are functions of cross-section per unit mass of DM particles, for example $\xi_1=\xi_1(\frac{\sigma}{m})$. In order to have our result consistent with M-T relation in a negligible $\frac{\sigma}{m}$, e.g. CDM, interacting constants should reduce to zero for $\frac{\sigma}{m}=0$. In other words, a larger value of $\frac{\sigma}{m}$ indicates more strong interaction between DE and DM and therefore, results in larger values of interacting constants. This would be the main effect of adding SIDM to the theory of interacting dark sector if future observations exactly result in CDM predictions of cluster mass and temperature. In this regard, upcoming surveys such as eROSITA (for measurement of clusters x-ray temperature) and LSST (for finding cluster masses via gravitational lensing) would play an impactful role in order to have a more precise value of coefficient factor in M-T relation. \\ 

To provide an opportunity for a more general conclusion, we iterated all steps of our approach for three other values of $\frac{\sigma}{m}$; they are $\frac{\sigma}{m}=0.1 \; \rm{cm}^2 g^{-1}$ (or SIDS0.1), $\frac{\sigma}{m}=0.4 \; \rm{cm}^2 g^{-1}$ (or SIDS0.4) and $\frac{\sigma}{m}=0.6 \; \rm{cm}^2 g^{-1}$ (or SIDS0.6). The only stage at which we lack information is velocity dispersion profile, since the SIDM simulations have not been performed for two of these cases. On the other hand, looking at Eqs.~(\ref{M_C_0.1}) and (\ref{M_C_1}) may help to iron out this issue. In the mass-velocity dispersion relation, power index is approximately $\approx \frac{1}{3}$, which is in agreement with the obtained relation in $\Lambda$CDM model. We fix the power index to this value and assume that the coefficient factor linearly changes from 1628 (for SIDM0.1) to 1876 (for SIDM1) with $\frac{\sigma}{m}$. Regarding this assumption, it is plausible to calculate $\lambda$ in a way that the M-T relation does not show any upward shift. As long as the quantity of $\lambda$ is determined, interacting constants for each model can immediately be found. Table~\ref{table_2} shows these obtained values of $\lambda$ and interacting constants for the given models. As there are two interacting constants for Model I, we separate this model into two particular cases of $\alpha_{DE}=0$ with nonzero $\alpha_{DM}$, and $\alpha_{DM}=0$ with nonzero $\alpha_{DE}$. \\

\begin{table}[h]
\caption{The calculated values of the virial ratio and interacting constants for SIDS with different values of $\frac{\sigma}{m}$.}
\begin{tabular}{c|c|c|c|c|c|c|}
\cline{2-7}
                              & $\lambda$ & \begin{tabular}[c]{@{}c@{}}$\alpha_{DM}$\\ ($\alpha_{DE}=0$)\end{tabular} & \begin{tabular}[c]{@{}c@{}}$\alpha_{DE}$\\ ($\alpha_{DM}=0$)\end{tabular} & $\xi_1$       & $\xi_2$       & $\xi_3$       \\ \hline
\multicolumn{1}{|c|}{CDM} & 0.500    & 0                                             & 0                                             & 0       & 0       & 0       \\ \hline
\multicolumn{1}{|c|}{SIDS0.1} & 0.509   & $-0.0024$                                       & $-0.0044$                                       & $-0.0033$ & $-0.0060$ & $-0.0088$ \\ \hline
\multicolumn{1}{|c|}{SIDS0.4} & 0.533   & $-0.0087$                                       & $-0.0153$                                       & $-0.0119$ & $-0.0210$ & $-0.0321$ \\ \hline
\multicolumn{1}{|c|}{SIDS0.6} & 0.543   & $-0.0113$                                       & $-0.0196$                                       & $-0.0155$ & $-0.0269$ & $-0.0416$ \\ \hline
\multicolumn{1}{|c|}{SIDS1}   & 0.550   & $-0.0131$                                       & $-0.0225$                                       & $-0.0179$ & $-0.0309$ & $-0.0483$ \\ \hline
\end{tabular}
\label{table_2}
\end{table}

\begin{figure}[h]
 	\centering
 	\includegraphics [width=0.7 \linewidth] {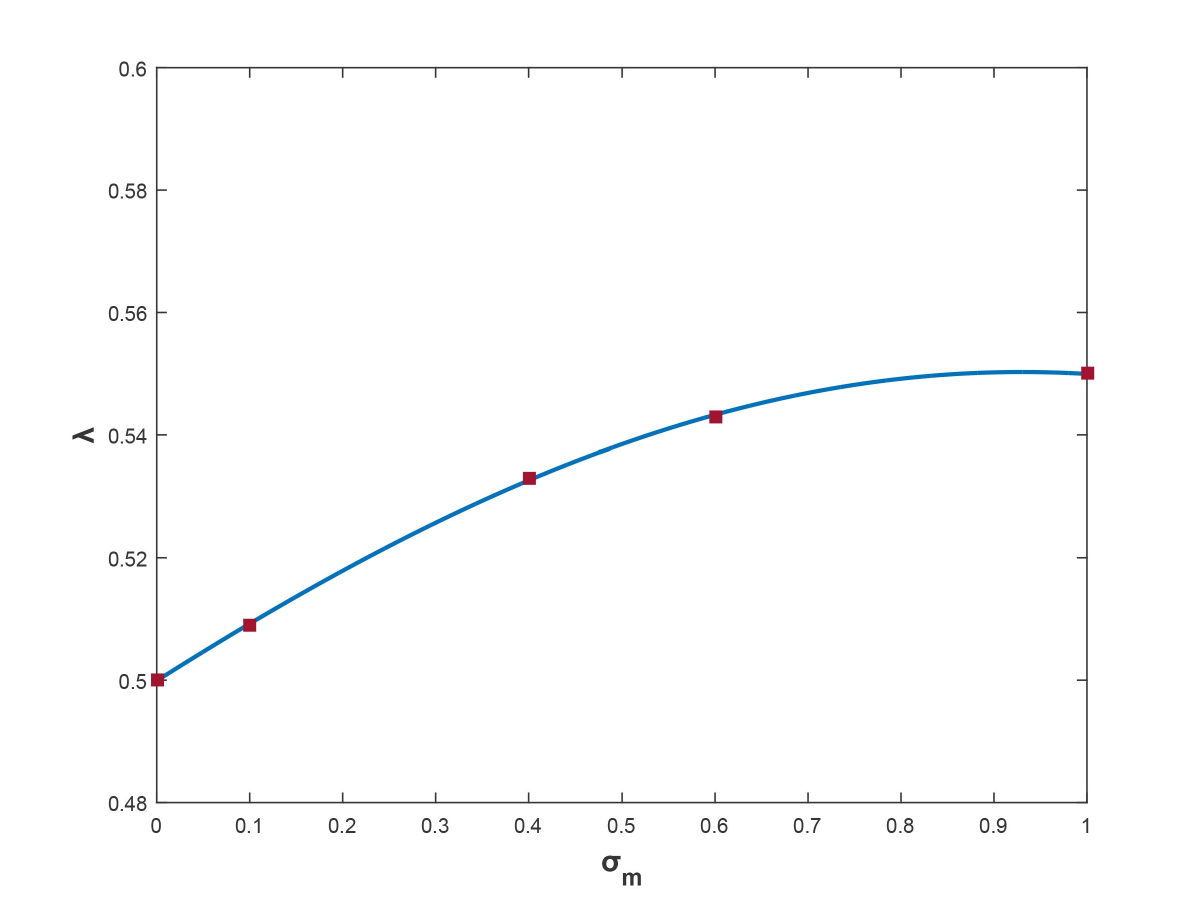}
 	\caption{The virial ratio as a function of $\sigma_m$; the squares are selected from Table~\ref{table_2} and the curve is drawn with respect to Eq.~(\ref{fourier}) with given constants.}
 	\label{figure_11}
 \end{figure}

According to the Table~\ref{table_2}, the virial ratio grows with $\frac{\sigma}{m}$ (or $\sigma_m$). Thus, $\lambda$ can be expressed as a function of $\sigma_m$ via the first terms of a Fourier series with the form of 

\begin{equation}\label{fourier}
\lambda = a_0 + a_1 cos(\omega \sigma_m) + b_1 sin(\omega \sigma_m)\, ,
\end{equation}

where $a_0=0.4649$, $a_1=0.03497$, $b_1=0.07784$ and $\omega = 1.235$. Fig. (\ref{figure_11}) indicates this function and values of $\lambda$ from Table~\ref{table_2}. With $RMSE=0.0005$ for this model, it is clear that Eq.~(\ref{fourier}) properly describes how the virial ratio changes with $\sigma_m$. \\

\begin{figure}[h]
 	\centering
 	\includegraphics [width=0.7 \linewidth] {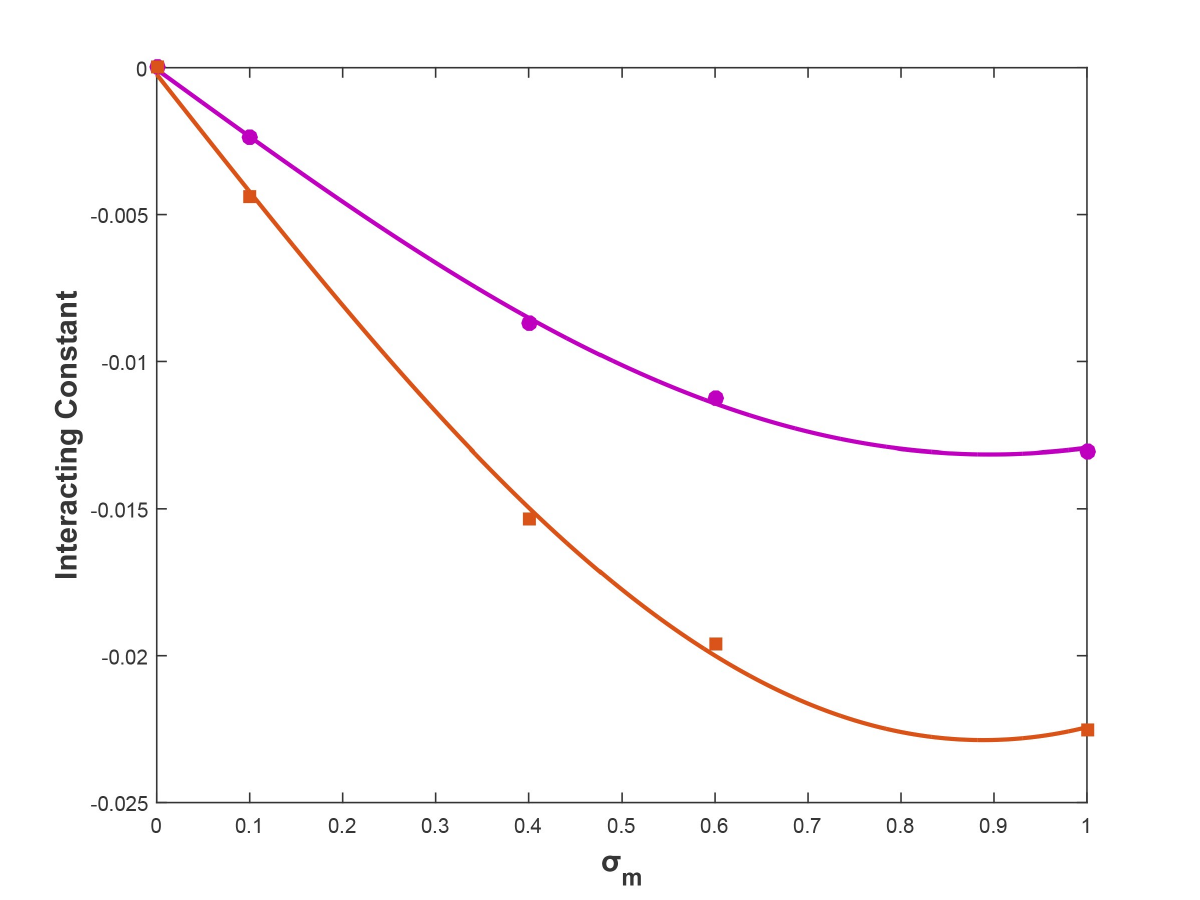}
 	\caption{Interacting constants of Model I as a function of $\sigma_m$ for two specific cases of $\alpha_{DE}=0$ (magenta) and $\alpha_{DM}=0$ (orange). In this plot, vertical axis reveals $\alpha_{DM}$ for magenta curve, and $\alpha_{DE}$ for orange curve. The points are shown according to Table~\ref{table_2} and curves are indicated regarding Eq.~(\ref{fitting_sine}) with the matched values of constants from Table~\ref{table_3}.}
 	\label{figure_12}
 \end{figure}

\begin{figure}[h]
 	\centering
 	\includegraphics [width=0.7 \linewidth] {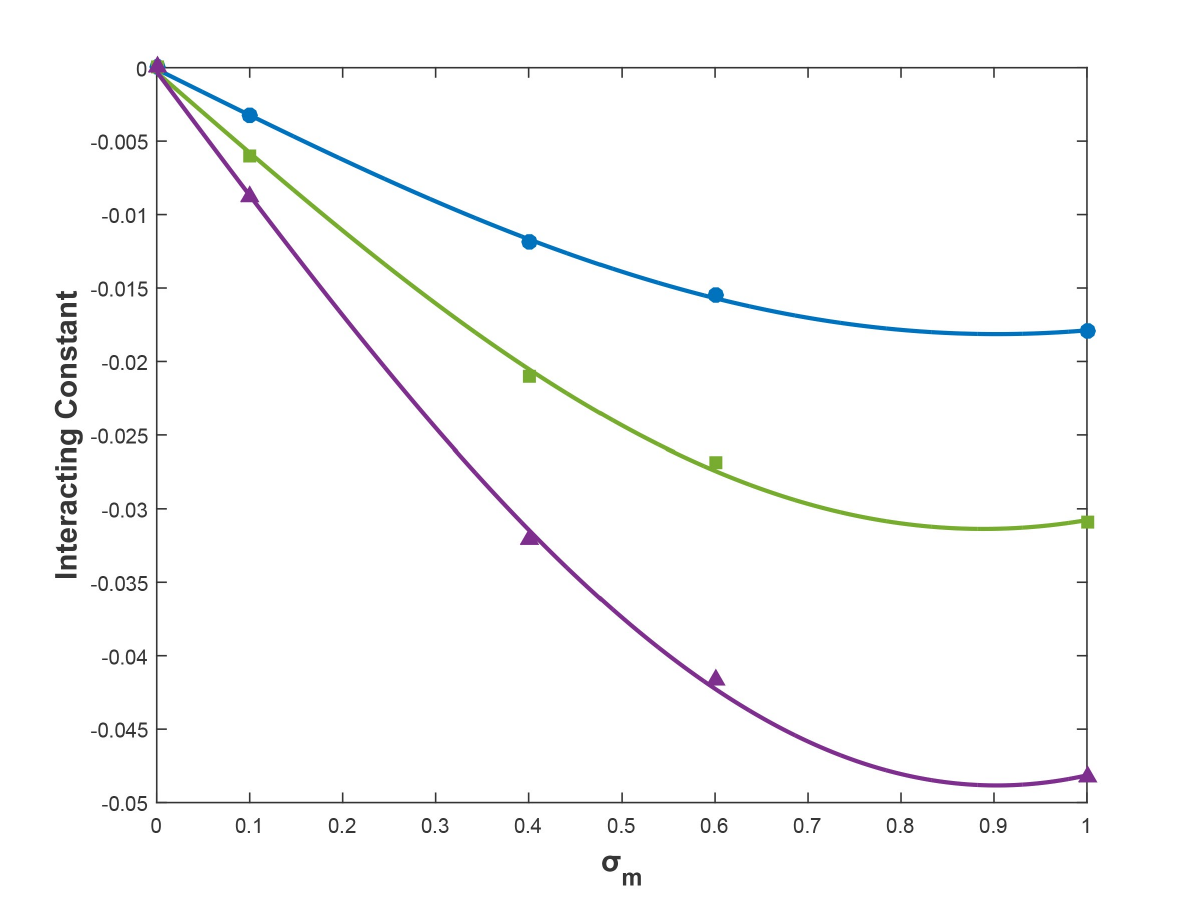}
 	\caption{The same plot as Fig. \ref{figure_12}, but with regard to Model II (blue), Model III (green) and Model IV (purple).}
 	\label{figure_13}
 \end{figure}

Finding interacting constants for various $\sigma_m$ values also results in a useful piece of information, which is the quantitative behavior of interacting constant of each model as a function of $\sigma_m$. Considering the relation of

\begin{equation}\label{fitting_sine}
\chi_{\rm{interaction}} = A_0 sin(\omega_0 \sigma_m + \phi)\, ,
\end{equation}

where the three constants of $A_0$, $\omega_0$ and $\phi$ can be determined via fitting this relation to the obtained constants in Table~\ref{table_2} for each model. In this expression, $\chi_{\rm{interaction}}=\alpha_{DM}, \alpha_{DE}, \xi_1, \xi_2, \xi_3$ denotes the corresponding interacting constant to each model. The result is collected in Table~\ref{table_3}. For two particular cases of Model I and for three other models, this result is shown in Figs. (\ref{figure_12}) and (\ref{figure_13}), respectively. These plots show that Eq.~(\ref{fitting_sine}) can be a true description for interacting constants as a function of $\sigma_m$. \\

The last conclusion has an specific importance. For inetracting dark sector models, $Q$ contains the quality of energy transfer between DE and DM. It is a function of DM and DE densities, in addition to several constants. While in interacting dark sector, these constants do not necessarily have a physical meaning, they directly describe DM particles characteristics in SIDS model. DM particles with higher $\frac{\sigma}{m}$ can transfer more substantial amount of energy to DE. In fact, the essence of interacting constants is specified in the SIDS model. \\

It is not necessary for all interacting dark sector models to obey Eq.~(\ref{Q_Models}) and to stay proportional to the Hubble parameter; these models may emerge in many different forms. For the other models of interacting dark sector (except for the ones introduced in Eq.~(\ref{Q_Models})), the same procedure is still valid to determine interacting constants. For every model, $\lambda$ of the virial condition is theoretically calculated via the method provided in Section II of Ref. \cite{Naseri:2020uvn}. Then, Fig. (\ref{figure_11}) can be used to find a relation between the interacting constant and $\sigma_m$. \\

Interacting constants can also be calculated with respect to the other cosmological tests. In the SIDS model, having a fixed value for these constants may help to constrain the key parameter of the collisional DM model, that is $\sigma_m$. In an observational perspective, in addition to finding constraints on the rate of energy transfer between DE and DM, more exact datasets of mass and temperature of galaxy clusters assist to test the scenario of SIDS model.

\begin{table}[h]
\caption{The obtained constants of Eq.~(\ref{fitting_sine}) with regard to the result of Table~\ref{table_2} for five models of SIDS.}
\begin{tabular}{c|c|c|c|c|}
\cline{2-5}
                                                                                & $A_0$       & $\omega_0$     & $\phi$  &$RMSE$     \\ \hline
\multicolumn{1}{|c|}{\begin{tabular}[c]{@{}c@{}}Model I\\ ($\alpha_{DE}=0$)\end{tabular}}  & 0.01315  & 1.749   & 3.146   &0.0002  \\ \hline
\multicolumn{1}{|c|}{\begin{tabular}[c]{@{}c@{}}Model I\\ ($\alpha_{DM}=0$)\end{tabular}} & 0.02287  & 1.757   & 3.152   &0.0004   \\ \hline
\multicolumn{1}{|c|}{Model II}                                                  & 0.01814 & 1.733 & 3.147   &0.0002  \\ \hline
\multicolumn{1}{|c|}{Model III}                                                 & 0.03137 & 1.757 & 3.152    &0.0006   \\ \hline
\multicolumn{1}{|c|}{Model IV}                                                  & 0.04883  & 1.733 & 3.147   &0.0007 \\ \hline
\end{tabular}
\label{table_3}
\end{table}

\section{Conclusion}
We studied halo profiles and scaling relations of galaxy clusters regarding SIDM halos. Based on the result of the previously performed simulations, we proposed a function which can perfectly describe the simulated density profile for the cross-section per unit mass of DM particles limited to $0.1 \; \rm{cm}^2 g^{-1}<\frac{\sigma}{m}<1 \; \rm{cm}^2 g^{-1}$ and cluster masses in the range of $10^{14}<\frac{\rm{M}}{\rm{M}_{\odot}}<10^{15}$. Unlike many other density profiles, this profile is not totally cored and is flexible to change with $\frac{\sigma}{m}$ and cluster mass, just as SIDM simulations predict. For this profile, a relation between cluster mass and the concentration parameter is found, which reveals higher values of concentration parameter in SIDM than the NFW profile for the same masses. \\

Solving the Jeans equation with several assumptions led to a straightforward but rather precise form of velocity dispersion profile. We also found the scaling relation between cluster mass and mean velocity dispersion inside the virial radius for SIDM0.1 and SIDM1. These results were used to find the impact of self-interaction of DM particles on the mass-temperature relation in galaxy clusters. It demonstrates that any value of observed x-ray temperature is attributed to more massive clusters in SIDM model in comparison with CDM. \\

Despite the fact that observations according to diverse methods of mass measurement vary in a comparable range with this upward shift, we find it beneficial to study a new hybrid interacting model, namely super interacting dark sector, which can effectively remove any difference in the coefficient factor of M-T relation. In this model, there is a plausible interaction between DM and DE, in addition to self-interaction among DM particles. Such a combination of two models of DM and DE can justify a number of cosmological problems and also cancel the effect of SIDM in mass-temperature relation. \\

Studying M-T relation suggested that constants of interacting models should change with $\frac{\sigma}{m}$. The higher the cross-section per unit mass of DM particles is, the stronger the interaction between DE and DM should be. The virial ratio is obtained to be higher than $\frac{1}{2}$ for SIDS models and accordingly, the interacting constants are found to be negative, which indicate that energy flows from DM to DE. \\

We found many quantitative expressions for the virial ratio and interacting constants as functions of $\sigma_m$. It is shown that in SIDS model, interacting constants have a physical meaning, which is related to DM particles mass and cross-section of self-interaction. A higher value of $\frac{\sigma}{m}$ reveals a more considerable amount of energy transfer from DM particles to DE. \\

However, if the future observations approve such a minute upward shift in M-T relation, SIDM effects can exclusively justify the result and there would be no need for SIDS model. In fact, the provided method in this study can be used for any observational dataset to investigate the constants of SIDS model. \\

The approach of this research is based on the outcomes of previous SIDM simulations. Any development in these simulations can help to improve our method and find better assumptions for a more precise study of both SIDS model and scaling relations of SIDM. As a result, the provided results will improve if the more intricate and comprehensive simulations are performed. In particular, a wide range of baryonic effects would make a notable contribution to the results of these simulations. \\

\appendix
\section{Statistical Analysis}
There are various parameters to evaluate how a fitted model argees with a dataset. One of the most common statistical parameters is "Root Mean Square Error", abbreviated to $RMSE$. It is defined as
\begin{equation}\label{RMSE}
RMSE=\sqrt{\frac{\sum\limits_{i=1}^{n} (y_i - \hat{y}_i)^2}{n}}\, ,
\end{equation}
where $y_i$ stands for $i$th element of the dataset, $\hat{y}_i$ is the predicted value of the model for the same element, and $n$ is the number of elements in the dataset. $RMSE$ has the same dimension as $y_i$ and ranges from zero to infinity. The lower value of $RMSE$ indicates a better fit between data and model. Clearly, $RMSE=0$ means that the model perfectly fits the data.

\section{Simulation Results}
In order to study the modified density profile for SIDM halos, we used the simulation results from Ref. \cite{Robertson:2018anx} and founded Eq.~(\ref{density_profile}) by fitting it to the data. In addition to SIDM effects, this simulation includes baryonic physics such as star formation, stellar evolution,  and stellar and AGN feedback. Undoubtedly, taking baryonic physics into account (i.e. full physics of SIDM halos) leads to a more realistic model. The simulation was performed for approximately 1000 halos with the masses of $13.9<\log(\frac{\rm{M}}{\rm{M}_{\odot}})<14.1$, and about 40 halos with $14.8<\log(\frac{\rm{M}}{\rm{M}_{\odot}})<15.2$. \\

\begin{figure}[h]
 	\centering
 	\includegraphics [width=0.7 \linewidth] {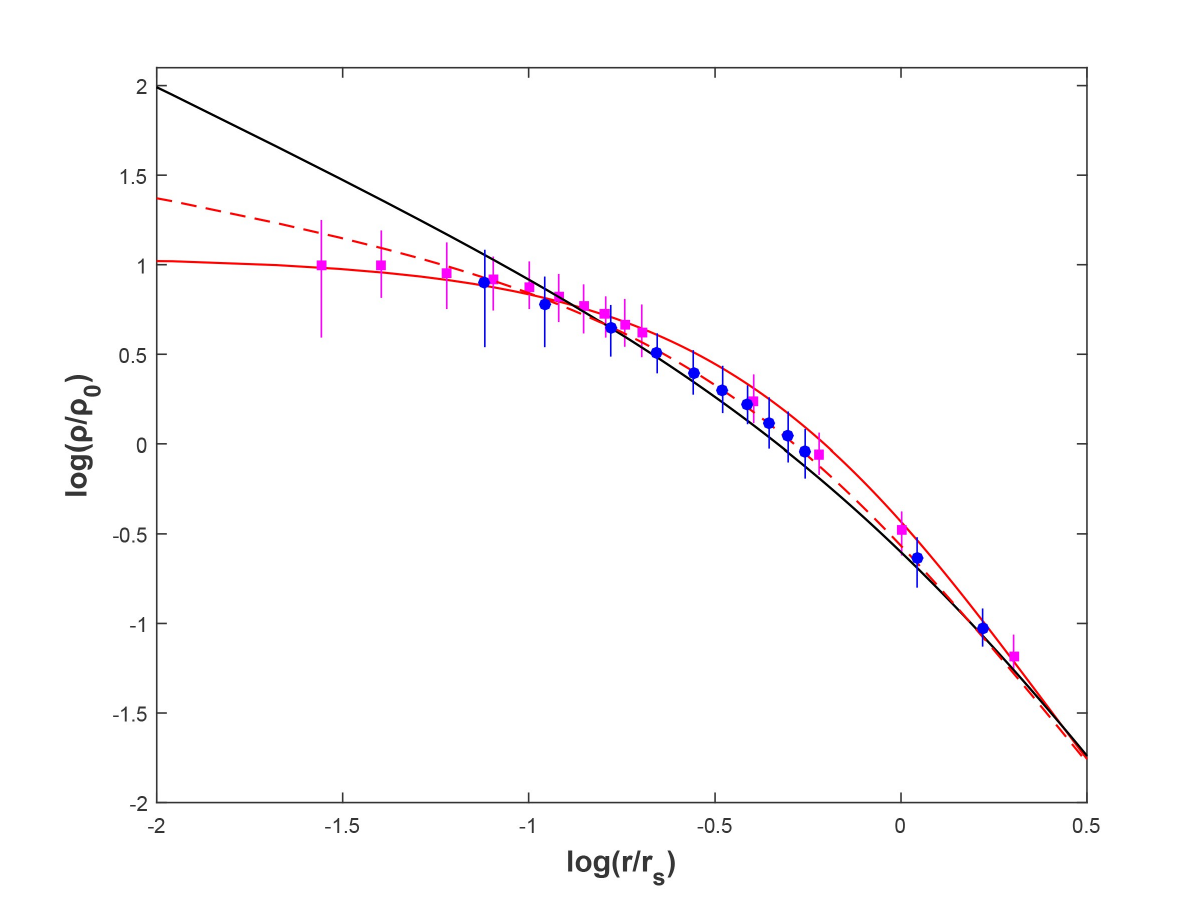}
 	\caption{A comparison between Eq.~(\ref{density_profile}) and simulation data for SIDM1 model. Red solid curve shows Eq.~(\ref{density_profile}) for $\frac{\rm{M}}{ \rm{M}_{\odot}} = 10^{15}$ (simulation data for this mass is illustreated by magenta color), while red dashed curve denotes this equation for $\frac{\rm{M}}{ \rm{M}_{\odot}} = 10^{14}$ (blue circles are the data from the simulation for this halo mass). The NFW profile is represented by black curve.}
 	\label{Appendix}
 \end{figure}

Fig. (\ref{Appendix}) illustrates the quality of fitting Eq.~(\ref{density_profile}) with the given parameters in Section II to the simulation. In this figure, we considered $A=0.2$, which has a better agreement with the data. Vertical lines that are matched to the data show 16-84th percentile ranges of the profile from the SIDM1 full physics simulation (the shaded region in Fig. 2 of the Ref. \cite{Robertson:2018anx}). According to the simulation details, for clusters with the masses of $\frac{\rm{M}}{ \rm{M}_{\odot}} \approx 10^{14}$ and $\frac{\rm{M}}{ \rm{M}_{\odot}} \approx 10^{15}$, we have $r_{200} \approx 960 \; \rm{kpc}$ and $r_{200} \approx 2080 \; \rm{kpc}$, respectively. We used the mass-concentration relation to calculate $r_s$ by Eq.~(\ref{concentration}), and then, transformed the radii given in the Ref. \cite{Robertson:2018anx} to our plots. It is clear that Eq.~(\ref{density_profile}) describes the simulation data with a good precision, as the statistical parameter of $RMSE$ for $\frac{\rm{M}}{ \rm{M}_{\odot}} = 10^{14}$ is $RMSE=0.0060$, and it is $RMSE=0.0109$ for $\frac{\rm{M}}{ \rm{M}_{\odot}} = 10^{15}$.

\end{document}